\documentclass[final,3p,times,twocolumn]{elsarticle}

\usepackage{amssymb}
\usepackage{amsmath}
\usepackage{amsfonts}
\usepackage[ruled,vlined]{algorithm2e}
\usepackage[caption=false,font=small,labelfont=small,textfont=small]{subfig}
\usepackage{tikz}
\usepackage{etoolbox}
\usepackage{wrapfig}
\usepackage{graphicx}
\usepackage{url}
\usepackage{hyperref}
\usepackage{pifont}
\usepackage{colortbl}

\journal{Future Generation Computer Systems}

\begin{document}

\begin{frontmatter}

\title{Instant Resonance: Dual Strategy Enhances the Data Consensus Success Rate of Blockchain Threshold Signature Oracles}

\author[label1,label2]{Youquan Xian}
\ead{xianyouquan@stu.gxnu.edu.cn}
\author[label1,label2]{Xueying Zeng}
\author[label1,label2]{Chunpei Li}
\author[label1,label2]{Dongcheng Li}
\author[label1,label2]{Peng Wang}
\author[label1,label2]{Peng Liu\corref{corresponding}}
\ead{liupeng@gxnu.edu.cn}
\author[label1,label2]{Xianxian Li\corref{corresponding}}
\ead{lixx@gxnu.edu.cn}

\affiliation[label1]{organization={Key Lab of Education Blockchain and Intelligent Technology, Ministry of Education},
            addressline={Guangxi Normal University},
            city={Guilin},
            postcode={541004},
            country={China}}

\affiliation[label2]{organization={School of Computer Science and Engineering},
            addressline={Guangxi Normal University},
            city={Guilin},
            postcode={541004},
            country={China}}

\cortext[corresponding]{Corresponding author.}

\begin{abstract}
With the rapid development of Decentralized Finance (DeFi) and Real-World Assets (RWA), the importance of blockchain oracles in real-time data acquisition has become increasingly prominent. Using cryptographic techniques, threshold signature oracles can achieve consensus on data from multiple nodes and provide corresponding proofs to ensure the credibility and security of the information. However, in real-time data acquisition, threshold signature methods face challenges such as data inconsistency and low success rates in heterogeneous environments, which limit their practical application potential. To address these issues, this paper proposes an innovative dual-strategy approach to enhance the success rate of data consensus in blockchain threshold signature oracles. Firstly, we introduce a Representative Enhanced Aggregation Strategy (REP-AG) that improves the representativeness of data submitted by nodes, ensuring consistency with data from other nodes, and thereby enhancing the usability of threshold signatures. Additionally, we present a Timing Optimization Strategy (TIM-OPT) that dynamically adjusts the timing of nodes' access to data sources to maximize consensus success rates. Experimental results indicate that REP-AG improves the aggregation success rate by approximately 56.6\% compared to the optimal baseline, while the implementation of TIM-OPT leads to an average increase of approximately 32.9\% in consensus success rates across all scenarios.

\end{abstract}



\end{frontmatter}


\section{Introduction}
As a core component of blockchain data interoperability, blockchain oracles play a crucial role in the acquisition and transmission of off-chain data, significantly advancing the development of blockchain applications \cite{pasdar2023connect, gigli2023decentralized}. In recent years, the rapid growth of fields such as Decentralized Finance (DeFi) \cite{deng2024safeguarding, kitzler2023disentangling} and Real World Assets (RWA) \cite{notheisen2017trading, hou2023chain} has led to increasing demand for external real-time data, such as exchange rates and price information \cite{zhao2022toward}.
Furthermore, other blockchain applications, including supply chain management \cite{powell2022garbage, manoj2023trusted, lee2023sensor}, the Internet of Things (IoT) \cite{shi2021blockchain, xian_iot, xian_iiot}, and smart cities \cite{esposito2021blockchain}, also rely on real-time data such as location and traffic flow to improve efficiency, transparency, and collaboration among stakeholders. In this context, oracles provide accurate and timely data, laying the foundation for the seamless integration of smart contracts in heterogeneous data environments while significantly enhancing the reliability and functionality of blockchain applications.

To ensure data credibility, blockchain oracles are typically composed of multiple distributed nodes that gather information from various data sources. Initially, the nodes collect data from multiple sources and perform preliminary aggregation using common methods such as median \cite{SchellingCoin, Oracul,tellor, Compound}, majority voting \cite{dong2023daon}, and weighted averaging \cite{xiao2023decentralized, xian2024safeguarding}. These aggregation techniques aim to improve the representativeness of the data and mitigate the influence of malicious data sources. Subsequently, the distributed nodes engage in a further consensus on the aggregated data to ensure the reliability of the result.
In this process, threshold signatures serve as a robust data consensus mechanism, ensuring that a valid signature is generated only when a predetermined number of nodes reach agreement. It effectively prevents the freeloading problem and provides strong cryptographic security guarantees, allowing the consensus results to be reliably verified. Consequently, threshold signatures have been widely adopted in various commercial projects, such as Chainlink \cite{chainlink} and the DOS Network \cite{dos}, becoming a critical technology for ensuring data credibility and security.

\begin{figure}[t]
    \centering
    \includegraphics[width=\linewidth]{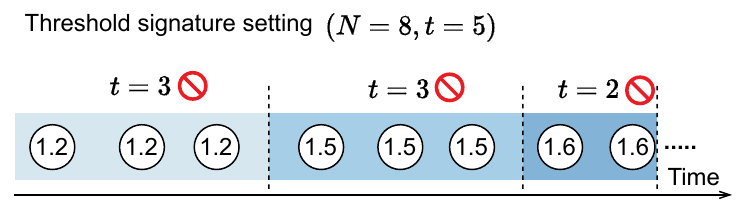}
    \caption{Threshold signature fails consensus when obtaining real-time data.}
    \label{fig:question}
\end{figure}

However, the timing of data retrieval by distributed heterogeneous nodes from distributed data sources is inconsistent, leading to heterogeneity in the acquired real-time data, which complicates the consensus requirements of threshold signatures. As illustrated in Figure \ref{fig:question}, different nodes may access information from the same data source at different times, resulting in discrepancies in both the timestamps and content of the data. This inconsistency significantly reduces the success rate of consensus in threshold signatures, affecting their usability in real-time data acquisition applications. Therefore, improving the consistency of aggregated data among nodes in real-time data retrieval tasks to ensure the usability of threshold signatures is the primary research objective of this paper.

In this paper, we propose two innovative strategies to improve the data consensus success rate of blockchain threshold signature oracles. First, we introduce a novel data aggregation method called the Representative Enhanced Aggregation Strategy (REP-AG), which aims to improve the representativeness of the data aggregated by nodes, ensuring consistency with the aggregated data of other nodes and thereby enhancing the usability of threshold signatures. Second, we design a Timing Optimization Strategy (TIM-OPT) that adjusts the timing of node access to data sources and data distribution, thus increasing the success rate of data consensus. 

The main contributions of this paper are as follows:
\begin{itemize} 
\item We propose a novel data aggregation method, REP-AG, which models the data aggregation process of nodes under incomplete information as a Bayesian game and solves it. This approach significantly improves the consistency of aggregation results among nodes, thereby enhancing the success rate of data consensus. 
\item We design a Timing Optimization Strategy TIM-OPT that introduces an appropriate waiting time before nodes access data sources, utilizing Bayesian game methods. This strategy effectively increases the concentration of data among nodes, thereby improving the success rate of data consensus.  
\item Experiments show that under the same environmental assumptions, REP-AG improves the consensus success rate by approximately 56.6 \% compared with the optimal baseline, and the consensus success rate of all schemes after the application of TIM-OPT increases by approximately 32.9 \% on average.
\end{itemize}

The remainder of this paper is structured as follows: Section \ref{bg} introduces the related works and existing challenges. Section \ref{main} presents the system workflow and details of the proposed approach. Section \ref{experiment} provides the experimental results and analysis. Finally, Section \ref{conclusion} concludes the paper and outlines future research directions.

\section{Related Work and Existing Problems}
\label{bg}

To establish consensus on data sourced from multiple nodes, early oracles frequently utilized on-chain consensus mechanisms, including voting and median aggregation. For example, systems such as Augur \cite{peterson2015augur}, Astraea \cite{adler2018astraea}, and research conducted by Cai et al. \cite{cai2020truth, cai2022truthful} employed voting game frameworks to consolidate data submitted by nodes on the blockchain, with the data validated through majority voting serving as the definitive outcome. Furthermore, selecting the median from the data collected on-chain from nodes has also been a prevalent approach for achieving consensus \cite{SchellingCoin, Oracul, tellor, Compound}. 
Although on-chain consensus is relatively simple to implement, it necessitates the storage of all nodes' data on-chain, resulting in considerable resource inefficiency. As a result, off-chain consensus methodologies have attracted increasing interest in recent years \cite{BandarupalliBBKLR24, WooSP20, 10061682}. For instance, Gigli et al. \cite{gigli2023decentralized, gigli2024zonia} established consensus on data aggregated off-chain using the Truth Inference algorithm. Specifically, threshold signatures facilitate data consensus through advanced cryptographic methods, which not only bolster security and fault tolerance but also effectively address the issue of freeloading, leading to their extensive utilization in oracle systems. Manoj et al. \cite{TMN23} implemented threshold signature oracles in the context of agricultural production risk management to streamline the acquisition of data from agricultural IoT devices integrated within the blockchain framework. Moreover, Yu et al. \cite{yu2023lattice} enhanced threshold signature oracles to resist potential quantum computing threats. Additionally, threshold signatures are prominently utilized as a fundamental technology for off-chain consensus in prominent commercial initiatives such as Chainlink \cite{chainlink} and DOS Network \cite{dos}.

In addition to facing distrust from nodes, oracles also confront a trust crisis concerning the reliability of data sources. Consequently, nodes typically need to access multiple data sources and aggregate the information to mitigate the influence of malicious sources \cite{lv2021blockchain, almi2023graph, gigli2023decentralized}. If nodes return all data without aggregation, the risk of successful attacks by malicious nodes may increase, jeopardizing the security of the system. To address this issue, researchers have proposed various aggregation methods.
Beyond the traditional median-based approaches commonly used in early on-chain systems, researchers such as Xiao \cite{xiao2023decentralized} and Xian \cite{xian2024safeguarding} have employed Truth Discovery methods to perform weighted aggregation of data from different sources off-chain, based on the credibility of the data sources. This approach continually updates the reliability assessments, allowing the aggregated data to converge toward the truth. Additionally, DAON \cite{dong2023daon} explores the use of majority voting or averaging methods at the node level to diminish the impact of erroneous data on the final results of threshold signature aggregation. These methods provide effective solutions for enhancing the accuracy and reliability of data aggregation.

\subsection{Existing Problems}
\label{questions}

Despite the extensive body of existing research, real-time data acquisition tasks for oracles operating in heterogeneous environments continue to encounter two primary challenges:

\paragraph{\textbf{Inconsistent Aggregation Results}} In the realm of real-time data acquisition within heterogeneous settings, traditional data aggregation methodologies frequently struggle to maintain consistency among the data collected by various nodes. This lack of consistency can undermine the success rate of consensus during the threshold signature consensus phase.

\paragraph{\textbf{High Variability in Access Timing}} At present, there is an absence of effective strategies to address the inconsistencies in the timing of nodes accessing data sources in heterogeneous environments. This situation results in diminished data consistency and poses significant risks to the success rate of threshold signature consensus.

\section{The Proposed Scheme}
\label{main}

In this section, we will introduce the workflow of the proposed oracle solution, along with the implementation details of the proposed dual-strategy approach.

\begin{figure*}[t]
    \centering
    \includegraphics[width=0.8\linewidth]{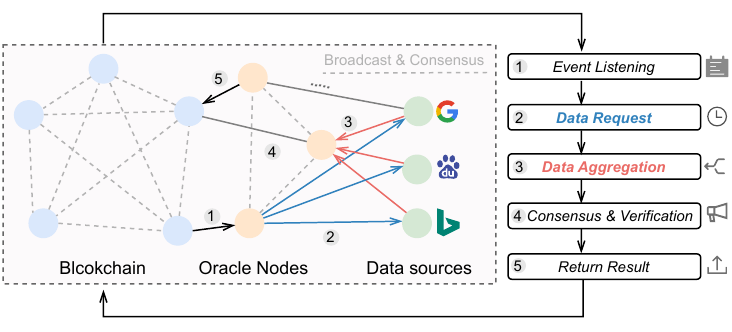}
    \caption{Process of the proposed solution.}
    \label{fig:network}
\end{figure*}

\subsection{System Flow}

The proposed threshold signature oracle system typically involves \( N \) nodes retrieving data from \( M \) data sources. Its core process, illustrated in Figure \ref{fig:network}, can be simplified into five steps: \ding{182} Event Listening; \ding{183} Data Request; \ding{184} Data Aggregation; \ding{185} Data Consensus and Validation; \ding{186} Return Result.

\paragraph{\ding{182} Event Listening} When an on-chain smart contract or a user invokes the oracle contract interface to initiate a data request event \( E \), the event includes a detailed description of the required data, such as the set of data sources \( \mathcal{D} = \{\mathcal{D}_1, \mathcal{D}_2, \ldots, \mathcal{D}_M\} \), which will be recorded on the blockchain. All oracle nodes \( \mathcal{O} \) continuously monitor for new events on the blockchain. If a relevant event is detected at time \( \mathcal{T}_i \), the nodes will immediately parse the event details and generate the corresponding data request task \( T(E) \).

\paragraph{\ding{183} Data Request} For the request task \( T(E) \), node \( \mathcal{O}_i \) will initiate a request \( Q_{i,j} \) to the specified data source \( \mathcal{D}_j \in \mathcal{D} \), with a duration of \( \overrightarrow{\mathcal{T}_{i,j}} \). Before this, the node will set a waiting time \( \widetilde{\mathcal{T}_{i,j}} \) based on the Time of Access Optimization Strategy (TIM-OPT). Consequently, for the request \( Q_{i,j} \), node \( \mathcal{O}_i \) will receive the data \( X_{i,j} \) at time \( \overline{\mathcal{T}_i} = \mathcal{T}_i + \overrightarrow{\mathcal{T}_{i,j}} + \widetilde{\mathcal{T}_{i,j}} \) (ignoring the node's own subtle computation response time). Ultimately, the node will aggregate the retrieved data \( X_{i,j} \) from the \( M \) data sources into a complete data set \( X_i = \{X_{i,1}, X_{i,2}, \ldots, X_{i,M}\} \). For more details on TIM-OPT, please refer to Section \ref{TIM-OPT}.

\paragraph{\ding{184} Data Aggregation} 
During the data aggregation phase, all nodes \( \mathcal{O} \) aggregate their comprehensive datasets \( X_i \) and identify a data representative \( R_i \) for consensus in the subsequent stages. Specifically, we utilize the designed REP-AG algorithm to select the data representative \( R_i \) from \( X_i \) that is most likely to facilitate successful aggregation, ensuring its similarity to the representatives chosen by other nodes, even in the absence of knowledge regarding their selections. For further details on REP-AG, please consult Section \ref{REP-AG}.

\paragraph{\ding{185} Data Consensus and Verification} In the process of achieving data consensus using threshold signatures, each node \( \mathcal{O}_i \) selects its data representative \( R_i \) from its dataset \( X_i \) and generates a partial signature \( \sigma_i = \text{Sign}(R_i, Sk_i) \) using its assigned sub-key \( Sk_i \). Node \( \mathcal{O}_i \) broadcasts its partial signature \( \sigma_i \) to the other nodes while waiting for their partial signatures. When a node collects a sufficient number of partial signatures \( \{\sigma_{1}, \sigma_{2}, \dots, \sigma_{t}\} \) that meet the threshold \( t \), it utilizes a threshold signature algorithm (such as BLS signatures \cite{boneh2004short}) to combine these partial signatures into a complete signature \( \sigma \). This complete signature can be verified on-chain using the group public key \( Pk \), without relying on the individual public keys of the nodes, thus demonstrating that the signature was generated by at least \( t \) nodes for the same data representative \( R^{*} \).

\paragraph{\ding{186} Return Result} Once the data verification is successful, the system invokes a callback interface to return the consensus result \( R^{*} \) to the smart contract or the requestor on the blockchain. The smart contract distributes rewards to the participating signing nodes \( \mathcal{O}_i \) based on the consensus result and the valid signatures.

\subsection{REP-AG}
\label{REP-AG}

The fundamental principle of executing data aggregation at the node level involves employing techniques such as majority voting or weighted averaging to reduce the influence of erroneous data originating from individual data sources. In this context, each node possesses knowledge solely of its own collected data set \( X_i = \{X_{i,1}, X_{i,2}, \dots, X_{i,M}\} \) and remains unaware of the overarching global data collection matrix \( \mathbf{X} \).

\begin{eqnarray}
    \mathbf{X} = \begin{bmatrix} 
X_{1,1} & X_{1,2} & \cdots & X_{1,M} \\
X_{2,1} & X_{2,2} & \cdots & X_{2,M} \\
\vdots & \vdots & \ddots & \vdots \\
X_{N,1} & X_{N,2} & \cdots & X_{N,M} \\
\end{bmatrix}
\end{eqnarray}

Existing approaches typically assume that the data returned from the same data source by different nodes is consistent. However, this assumption does not hold in real-time data retrieval tasks. In a heterogeneous environment, the timing \( \overline{\mathcal{T}_i} \) at which different nodes access the same data source \( \mathcal{D}_j \) may vary, and even a single node may access different data sources \( \mathcal{D}_j \) at different times, leading to discrepancies in the acquired data. For instance, it is possible that \( X_{1,1} \neq X_{2,1} \) and \( X_{1,1} \neq X_{1,2} \).

\begin{figure}[t]
    \centering
    \includegraphics[width=0.8\linewidth]{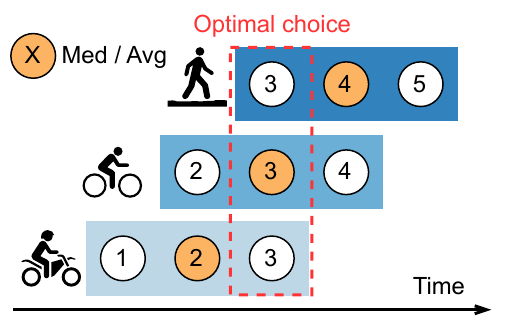}
    \caption{Optimal aggregation strategy in real-time data.}
    \label{fig:REP_AG}
\end{figure}

In this context, node \( \mathcal{O}_i \) employs conventional aggregation methods \( f_{\text{agg}} \), such as weighted averaging or median, to derive the aggregated data \( R_i \) from its data set \( X_i \). However, this aggregated result may not accurately represent the aggregation \( R_{\neg i} \) derived by other nodes from their respective data sets \( \{X_{j,1}, X_{j,2}, \dots, X_{j,M}\}_{j \ne i} \). As illustrated in Figure \ref{fig:REP_AG}, this discrepancy is the fundamental reason behind the challenge 1 encountered.

\begin{eqnarray}
    R_i = f_{\text{agg}}(X_{i,1}, X_{i,2}, \dots, X_{i,M})
\end{eqnarray}

To address this challenge, we propose the REP-AG aimed at improving the representativeness of the aggregated data from nodes under conditions of incomplete information. Since nodes lack complete knowledge of the decisions made by other nodes, we model the decision-making process of selecting \( R_i \) from \( X_i \) as a Bayesian game \cite{dekel2004learning,ely2005evolution}. This game can be formally represented as a quintuple \( G_1=(\mathcal{O}, A, \Theta, P, U) \).

\begin{itemize}
    \item Participant Set \(\mathcal{O}\): Represents the set of nodes participating in the game, i.e., \(\mathcal{O}=\{\mathcal{O}_1,\mathcal{O}_2,\dots,\mathcal{O}_N\}\), where \(N\) is the total number of nodes, and each \(\mathcal{O}_i\) represents an independent oracle node.
    
    \item Strategy Space \(A\): The set of strategies that each participant can choose, i.e., \(A = \{a_1,\dots,a_M\}\), where \(M\) is the total number of data sources, and \(a_j\) represents the selection of \(X_{i,j}\) as the data representative \(R_i\) from the data set \(X_i\).
    
    \item Type Set \(\Theta\): Contains all possible types \(\Theta = \{\theta_1, \theta_2, \ldots, \theta_M\}\), where each type \(\theta_j\) represents the preferences and advantages of node \(\mathcal{O}_i\) in selecting the data representative \(R_i\). Specifically, type \(\theta_j\) indicates the advantage of node \(\mathcal{O}_i\) choosing \(X_{i,j}\) as the aggregated data \(R_i\) compared to choosing other data \(X_{i,\neg j}\), reflecting the similarity of the selected \(R_i\) with the aggregated data of other nodes.
    
    \item Type Probability Distribution \(P\): Describes the prior probability distribution of each node \(\mathcal{O}_i\) over the types of other nodes. Specifically, \(P_{k,j}\) represents the probability that node \(\mathcal{O}_i\) believes node \(\mathcal{O}_k\) could belong to type \(\theta_j\). As the game progresses, nodes can dynamically update these beliefs by observing the behavior of other nodes, thereby improving the accuracy of their estimates regarding the types of other nodes.
    
    \item Utility function \( U \): Describes the payoff obtained by each node $\mathcal{O}_i$ based on its strategy selection when considering other node types $\theta_j$.
\end{itemize}

Given that nodes can only adjust their belief distribution gradually based on partial information during the game and cannot obtain a globally optimal strategy at once, we employ an iterative method to approach Nash equilibrium or a stable point of the system by updating the nodes' strategies and belief matrices \cite{quer2013inter}. The specific steps are outlined in Algorithm \ref{algorithm:1}, which can be divided into the following four steps: 1) Initialization; 2) Strategy Selection; 3) Observation and Update; 4) Repeat Iteration.

\begin{algorithm}[h]
\label{algorithm:1}
\caption{REP-AG}
\KwIn{Set of nodes $\mathcal{O}$, node data sets $\{X_i\}_{i \in N}$,  threshold $t$}

\tcp{Step 1}
\For{each node $\mathcal{O}_i \in \mathcal{O}$}{ 
    Initialize belief matrix $\mathbf{P}_i$\;
}  

\Repeat{belief matrix $\mathbf{P}_i$ and strategy $R_i$ converge}{
    \tcp{Step 2}
    \For{each node $\mathcal{O}_i \in \mathcal{O}$}{
        $\mathbb{E}[X_{i,j} | \mathbf{P}_i] \leftarrow \sum_{k \neq i} P_{k,j}$\;
        $\mathbb{R} \leftarrow \arg\max_{a_j \in A} \mathbb{E}[X_{i,j} | \mathbf{P}_i]$\;
    }
    \tcp{Step 3}
    $R^{*} \leftarrow ThresholdSignature(\mathbb{R}, t)$\;
    \For{each node $\mathcal{O}_i \in \mathcal{O}$}{
        \For{each result $R_k \in \mathbb{R}$}{
            \If{$R_k = R^{*}$}{
                $\Delta(R_k) \leftarrow 1$\; 
            }
            \Else{
                $\Delta(R_k) \leftarrow \frac{\sum_{R_i \in \mathbb{R}} \mathbb{I}(R_i, R_k)}{N}$\; 
            }
            \For{each result $X_{i,j} \in X_{i}$}{
                $P_{k,j} \leftarrow P_{k,j} + \Delta(R_k) \cdot \mathbb{I}(R_k, X_{i,j})$\;
            }
        }
    }
}
\end{algorithm}

\paragraph{Initialization} 
Each node \( \mathcal{O}_i \) randomly selects an initial strategy \( R_i \in A \). Simultaneously, each node sets an independent initial belief matrix \( \mathbf{P}_i \), where \( P_{k,j} \) represents the prior probability distribution of node \( \mathcal{O}_i \) regarding whether node \( \mathcal{O}_k \) belongs to type \( \theta_j \). Specifically, \( P_{k,j} \) indicates the preference degree of node \( \mathcal{O}_k \)’s aggregated result \( R_k \) after mapping to the strategy space \( X_i \) of node \( \mathcal{O}_i\), i.e., the probability that \( R_k \) is the same as \( X_{i,j} \).

\begin{eqnarray}
    \mathbf{P}_i = \begin{bmatrix} 
P_{1,1} & P_{1,2} & \cdots & P_{1,M} \\
P_{j,1} & P_{j,2} & \cdots & P_{j,M} \\
\vdots & \vdots & \ddots & \vdots \\
P_{N,1} & P_{N,2} & \cdots & P_{N,M} \\
\end{bmatrix}_{i \neq j}
\end{eqnarray}

\paragraph{Strategy Selection} 
In this game, the payoff for node \( \mathcal{O}_i \) is closely related to the aggregation results of other nodes \( R_{\neg i} \). For the threshold signature algorithm, the consensus is successful only if \( t \) nodes return the same aggregation result, thereby completing the task and earning a reward. Thus, the utility function \( U \) for the node can be expressed as:

\begin{eqnarray}
    U(a_j, R_{\neg i}) = 
\begin{cases} 
1, & \text{if } \sum_{k=1, k \ne i}^N \mathbb{I}(X_{i,j}, R_k) \geq t-1 \\ 
0, & \text{otherwise} 
\end{cases}
\end{eqnarray}

Here, the indicator function \( \mathbb{I}(X_{i,j}, R_k) \) returns 1 if \( X_{i,j} \) is the same as \( R_k \), and 0 otherwise.

Using the belief matrix \( \mathbf{P}_i \), node \( \mathcal{O}_i \) can estimate the expected aggregation result \( R_k \) for \( X_{i,j} \):

\begin{eqnarray}
\mathbb{E}[X_{i,j} | \mathbf{P}_i] = \sum_{k=1, k \ne i}^N P_{k,j} 
\end{eqnarray}

Based on the expected values from the belief matrix, node \( \mathcal{O}_i \) will select the strategy that maximizes its utility, which corresponds to maximizing the probability of selecting the same \( a^{*} \) as other nodes, thereby designating \( X_{i,j} \) as its data representative \( R_i \). Specifically, node \( \mathcal{O}_i \) will choose its strategy based on the estimated utility:

\begin{eqnarray}
    R_i \gets \arg\max_{a_j \in A} \mathbb{E}[X_{i,j} | \mathbf{P}_i]
\end{eqnarray}

\paragraph{Observation and Update} 
After each task round, the node will receive the broadcast result of the final consensus, \( R^{*} \). Additionally, all nodes share their aggregation results \( R_k \) after the consensus is completed (regardless of whether it was successful or failed), forming a shared result set \( \mathbb{R} \).

Each node checks whether the aggregation results \( R_k \) from other nodes can be mapped to its strategy \( a_j \) corresponding to the data \( X_{i,j} \). If so, it updates the belief at \( P_{k,j} \):

\begin{eqnarray}
    P_{k,j} \gets P_{k,j} + \Delta(R_k) \cdot \mathbb{I}(R_k, X_{i,j})
\end{eqnarray}

Here, \( \Delta \) represents the magnitude of belief enhancement. If \( R_k \) is the same as \( R^{*} \), the magnitude is 1. Otherwise, it depends on the number of results in \( \mathbb{R} \) that match \( R_k \):

\begin{eqnarray}
    \Delta(R_k) = 
\begin{cases} 
1, & \text{if } R_k = R^* \\ 
\frac{\sum_{R_i \in \mathbb{R}}{\mathbb{I}(R_i,R_k)} }{N}, & \text{otherwise} 
\end{cases}
\end{eqnarray}

\paragraph{Repeat Iteration} 
Continue executing steps 2 and 3, with each node gradually updating its belief matrix \( \mathbf{P}_i \) based on partially public information, thereby optimizing its belief distribution about other node types and continually improving the accuracy of its predictions regarding other node behaviors. The iterative process concludes when the belief matrix and strategies stabilize, reaching Nash equilibrium or an approximate equilibrium.

\subsection{TIM-OPT}
\label{TIM-OPT}

In a heterogeneous environment, the timing $\overline{\mathcal{T}_i}$ at which various nodes access the identical data source $\mathcal{D}_j$ can differ. Moreover, a single node may retrieve data from different sources $\mathcal{D}_j$ at distinct timing, resulting in the acquisition of varying data $X_{i,j}$. This phenomenon fundamentally underpins the emergence of Problem 2.

\begin{figure}[t]
    \centering
    \includegraphics[width=0.8\linewidth]{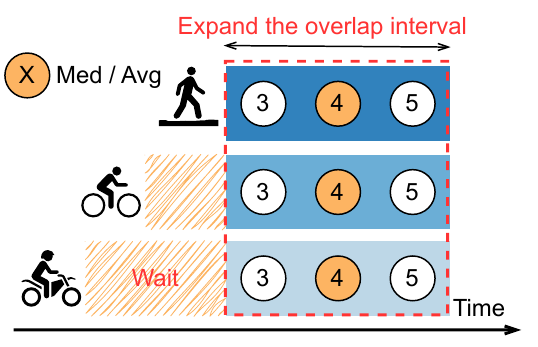}
    \caption{Sketch of TIM-OPT.}
    \label{fig:TIM_OPT}
\end{figure}

To address Problem 2, we design a timing optimization strategy (TIM-OPT) that adjusts the timing at which nodes access data sources, increasing the overlap of data and thereby maximizing the success rate of data consensus, as illustrated in Figure \ref{fig:TIM_OPT}. Similar to the design approach of REP-AG, TIM-OPT is also constructed as a Bayesian game in the form of a five-tuple $G_2=(\mathcal{O},A,\Theta,P,U)$.

The difference lies in that the strategy space \(A\) of \(G_2\) represents the waiting time \(\widetilde{\mathcal{T}_{i,j}}\) that node \(\mathcal{O}_i\) should pause before accessing \(\mathcal{D}_j\). To simplify the solution, we set this to a time interval \(\mathcal{T} = \{\frac{\Omega_i}{\Bbbk}, \frac{2 \cdot \Omega_i}{\Bbbk}, \dots, \Omega_i\}\) divided into \(\Bbbk\) equal parts, where \(\Omega_i\) represents the historical average access time interval length of node \(\mathcal{O}_i\).

The type set \(\Theta\) is given by \(\Theta = \{\theta_1, \theta_2, \ldots, \theta_n\}\), where type \(\theta_j\) indicates node \(\mathcal{O}_i\)'s preference for the waiting time when accessing data source \(\mathcal{D}_j\).

For the type probability distribution matrix \(\mathbf{P}_i\) of node \(\mathcal{O}_i\), \(P_{k,j,l}\) represents the probability that node \(\mathcal{O}_k\) is of type \(\theta_l\) regarding data source \(\mathcal{D}_j\).

\begin{eqnarray}
    \mathbf{P}_i = \{ P_{i,j,k} \mid i \in N; \, j \in M; \, k \in \Bbbk \}
\end{eqnarray}

The solving process is shown in Algorithm \ref{algorithm:2}. For node \(\mathcal{O}_i\) choosing a strategy to access data source \(\mathcal{D}_j\), the node first estimates the access strategies of other nodes regarding that data source based on the belief matrix \(\mathbf{P}_i\). Specifically, the node calculates the expected utility for strategy \(a_l\):

\begin{eqnarray}
\mathbb{E}[a_l | \mathbf{P}_i] = \sum_{k=1, k \ne i}^N P_{k,j,l} 
\end{eqnarray}

Next, node \(\mathcal{O}_i\) will choose the strategy \(a_l\) that maximizes the expected utility and set \(\mathcal{T}_l\) as its waiting time \(\widetilde{\mathcal{T}_{i,j}}\) before accessing data source \(\mathcal{D}_j\):

\begin{eqnarray}
    \mathcal{T}_l \gets \arg\max_{a_l \in A} \mathbb{E}[a_l | \mathbf{P}_i]
\end{eqnarray}

After each task concludes, node \(\mathcal{O}_i\) updates its belief matrix \(\mathbf{P}_i\) based on the shared result set \(\mathbb{R}\) and the consensus success result \(R^{*}\) (if it exists).

First, the node will traverse the shared result set \(\mathbb{R}\) and map \(R_k\) onto its strategy space \(A\) by checking whether \(R_k\) is the same as \(X_{i,j}\). The updated formula is as follows:

\begin{eqnarray}
    P_{k,j,l} \gets P_{k,j,l} + \Delta(R_k) \cdot \mathbb{I}(R_k, X_{i,j})
\end{eqnarray}

Here, \(\Delta(R_k)\) represents the magnitude of belief adjustment, which is calculated as follows:

\begin{eqnarray}
    \Delta(R_k) = 
\begin{cases} 
1, & \text{if } R_k = R^* \\ 
\frac{\sum_{R_i \in \mathbb{R}} \mathbb{I}(R_i, R_k) - t}{t}, & \text{otherwise} 
\end{cases}
\end{eqnarray}

In this process, if the aggregated result \(R_k\) matches the final consensus success result \(R^*\), then the belief adjustment magnitude is 1, indicating that the data \(X_{i,j}\) obtained by this strategy \(a_l\) is similar to the majority of results. It is important to note that if \(R_k\) differs from \(R^*\), a negative adjustment is set based on the number of results in the shared result set \(\mathbb{R}\) that match \(R_k\), to avoid the Matthew effect during the strategy selection process \cite{merton1968matthew}.

\begin{algorithm}[h]
\label{algorithm:2}
\caption{TIM-OPT}
\KwIn{Set of nodes $\mathcal{O}$, data set of nodes $\{X_i\}_{i \in N}$, threshold $t$}

\tcp{Step 1}
\For{each node $\mathcal{O}_i \in \mathcal{O}$}{ 
    Initialize belief matrix $\mathbf{P}_i$\;
}  

\Repeat{belief matrix $\mathbf{P}_i$ and strategy $R_i$ converge}{
    \tcp{Step 2}
    \For{each node $\mathcal{O}_i \in \mathcal{O}$}{
        \For{each data source $\mathcal{D}_j \in \mathcal{D}$}{
            $\mathbb{E}[a_l | \mathbf{P}_i] = \sum_{k=1, k \ne i}^N P_{k,j,l} $\;
            $\mathcal{T}_l \leftarrow \arg\max_{a_l \in A} \mathbb{E}[a_l | \mathbf{P}_i]$\;
            $Wait(\mathcal{T}_l)$\;
            $X_i \leftarrow RequestData(\mathcal{D}_j)$\;
        }
        $\mathbb{R} \leftarrow DataAggregation(X_i)$
    }
    
    \tcp{Step 3}
    $R^{*} \leftarrow ThresholdSignature(\mathbb{R}, t)$\;
    \For{each node $\mathcal{O}_i \in \mathcal{O}$}{
        \For{each result $R_k \in \mathbb{R}$}{
            \If{$R_k = R^{*}$}{
                $\Delta(R_k) \leftarrow 1$\; 
            }
            \Else{
                $\Delta(R_k) \leftarrow \frac{\sum_{R_i \in \mathbb{R}}{\mathbb{I}(R_i, R_k)} - t}{t}$\; 
            }
            \For{each result $X_{i,j} \in X_{i}$}{
                $P_{k,j,l} \gets P_{k,j,l} + \Delta(R_k) \cdot \mathbb{I}(R_k, X_{i,j})$\;
            }
        }
    }
}
\end{algorithm}

\begin{table}[h]
\centering
\caption{Experimental parameter settings.}
\resizebox{\linewidth}{!}{%
\begin{tabular}{lll}
\hline
Parameter & Value \\ \hline
Number of nodes \(N\) & 21 \\
Number of data sources \(M\) & 5 \\
Threshold \(t\) & 11 \\
Number of tasks & 1000 \\
Gaussian network latency distribution & \(\max(0, N(0.7, 0.2887))\) \\
Uniform network latency distribution & \(\max(0, U(0.02, 1.02))\) \\
Random perturbation distribution & \(U(0, 0.1)\) \\
Data source state change frequency \(f\) & 5 Hz \\
Size of TIM-OPT strategy space \(\Bbbk\) & 10 \\
\hline
\end{tabular}
}
\label{tab:setting}
\end{table}

\section{Experiment and Result Analysis}
\label{experiment}

 \begin{figure*}[t]
\centering
\subfloat[$f$=2 Hz]{\includegraphics[width=0.22\linewidth]{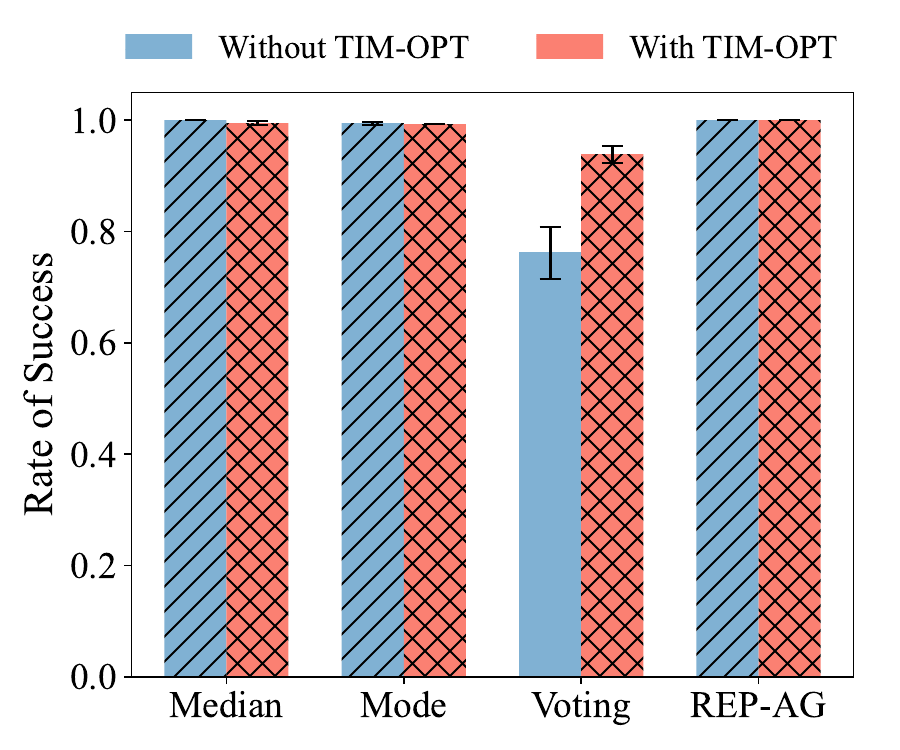}%
\label{fig:all_gaussian_2}}
\hfil
\subfloat[$f$=5 Hz]{\includegraphics[width=0.22\linewidth]{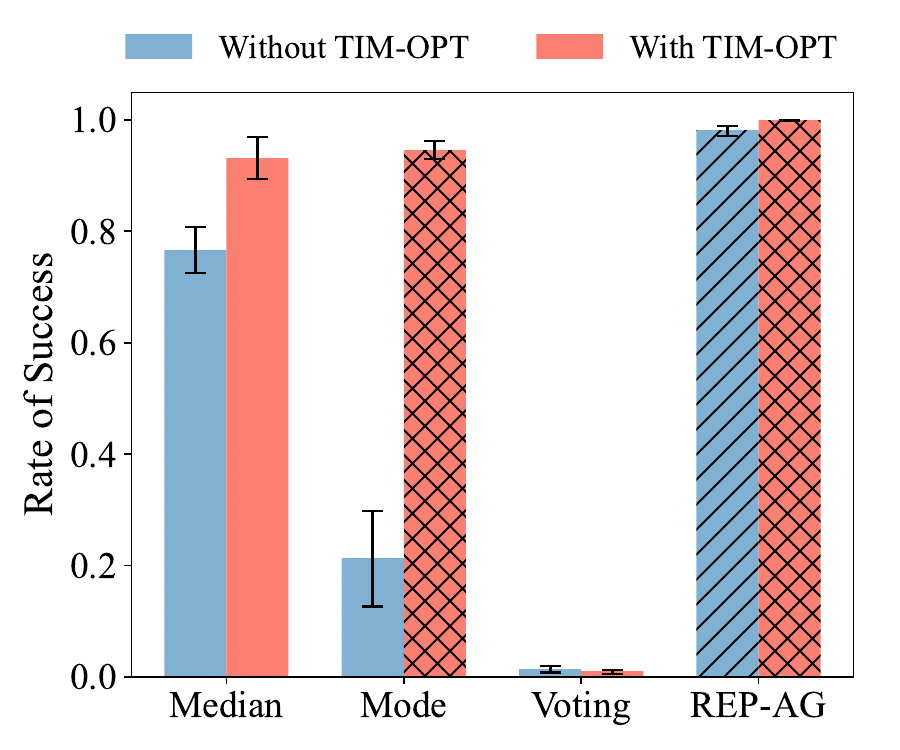}%
\label{fig:all_gaussian_5}}
\hfil
\subfloat[$f$=8 Hz]{\includegraphics[width=0.22\linewidth]{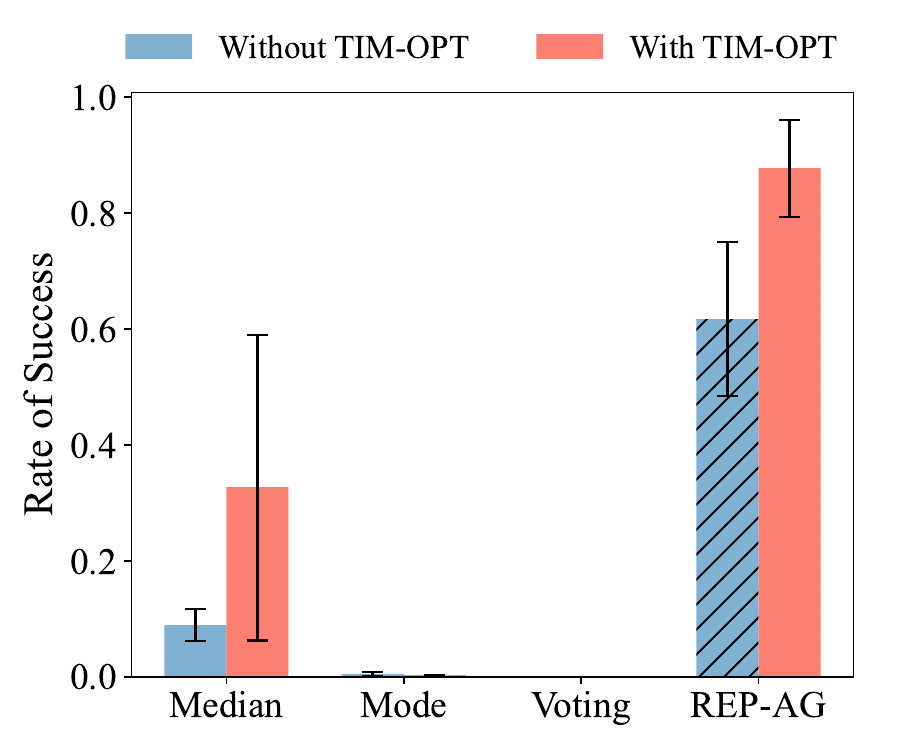}%
\label{fig:all_gaussian_8}}
\hfil
\subfloat[$f$=10 Hz]{\includegraphics[width=0.22\linewidth]{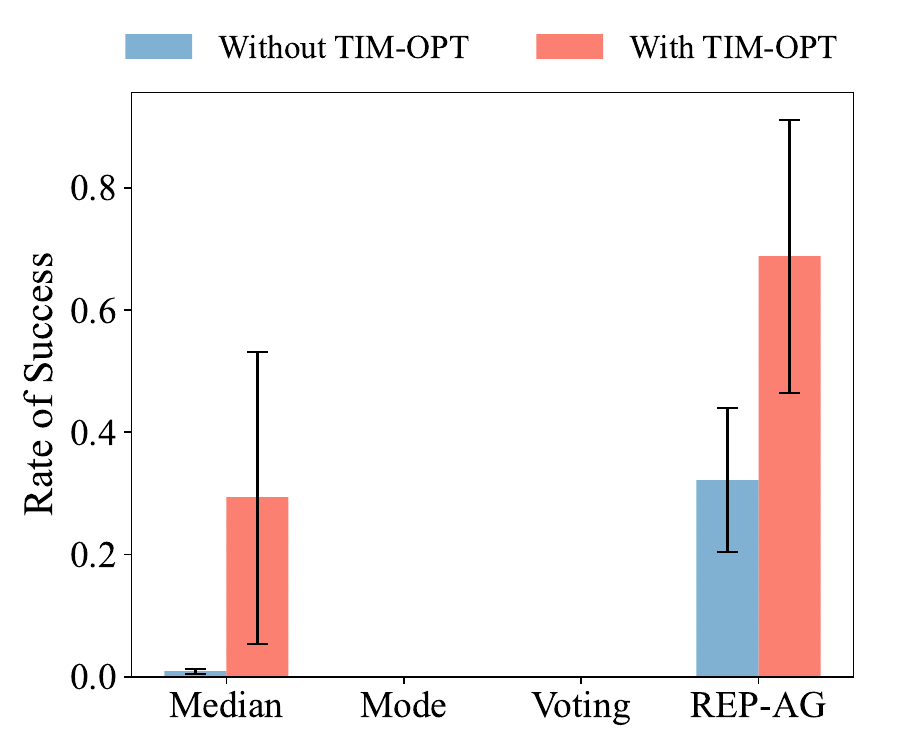}%
\label{fig:all_gaussian_10}}
\caption{Gaussian network latency distribution.}
\label{fig:all_gaussian}
\end{figure*}

 \begin{figure*}[h]
\centering
\subfloat[$f$=2 Hz]{\includegraphics[width=0.22\linewidth]{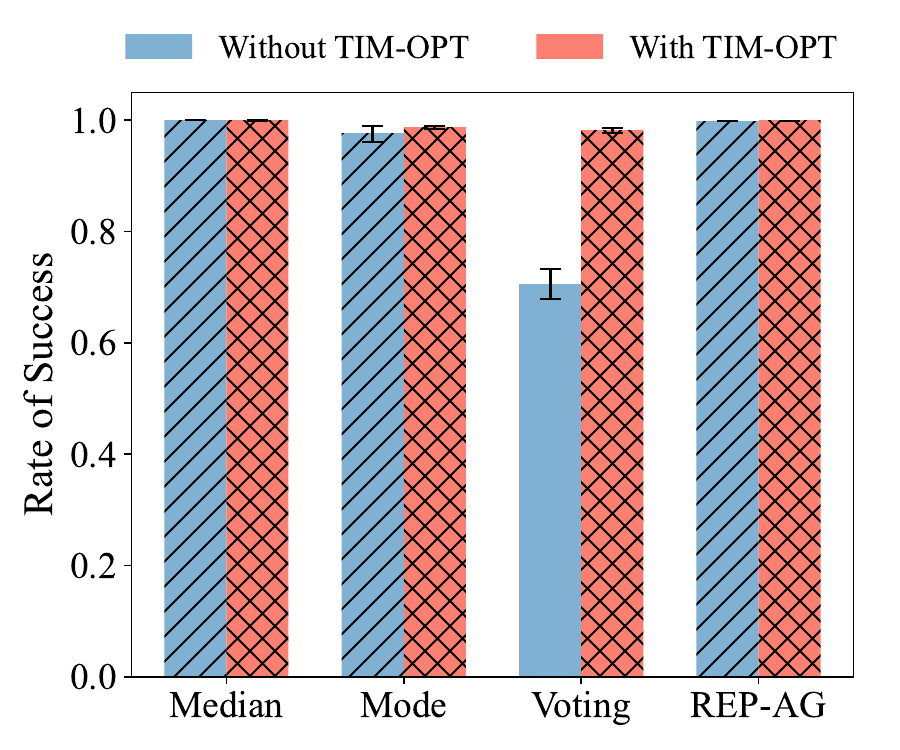}%
\label{fig:all_random_2}}
\hfil
\subfloat[$f$=5 Hz]{\includegraphics[width=0.22\linewidth]{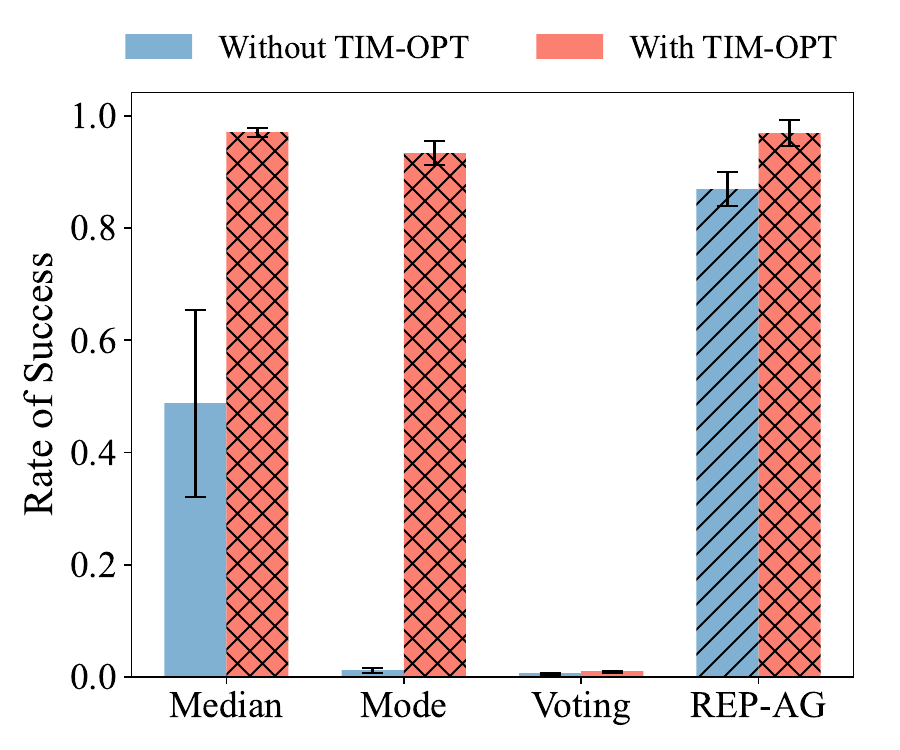}%
\label{fig:all_random_5}}
\hfil
\subfloat[$f$=8 Hz]{\includegraphics[width=0.22\linewidth]{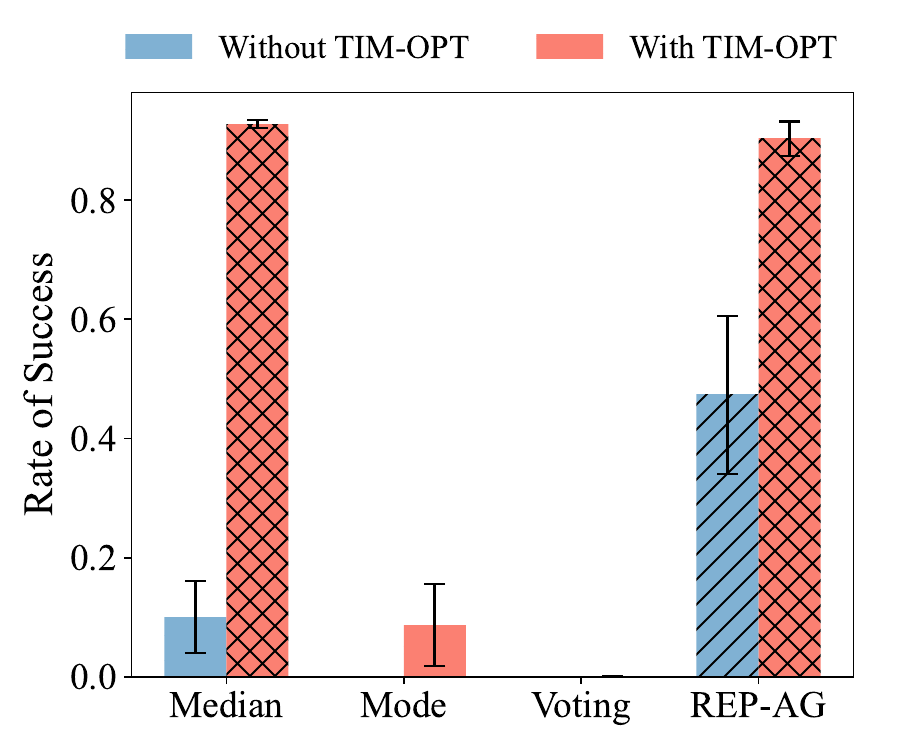}%
\label{fig:all_random_8}}
\hfil
\subfloat[$f$=10 Hz]{\includegraphics[width=0.22\linewidth]{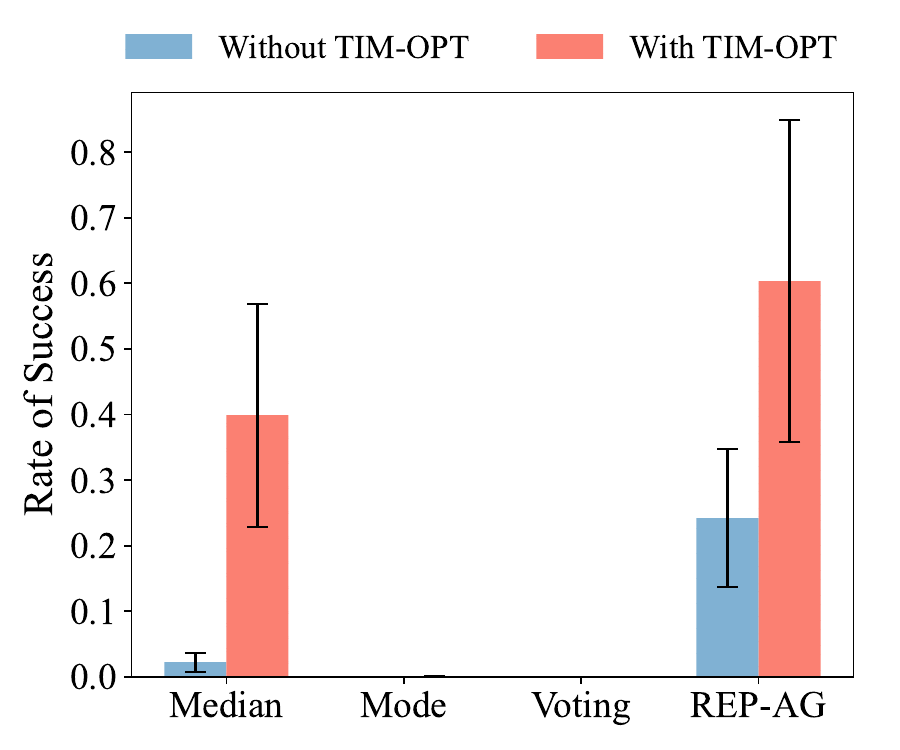}%
\label{fig:all_random_10}}
\caption{Uniform network latency distribution.}
\label{fig:all_random}
\end{figure*}

To validate the effectiveness of the proposed schemes (REP-AG and TIM-OPT), we implemented a prototype of these schemes and conducted tests in a controlled experimental environment. The hardware configuration for the experiments included an 8-core Intel i7-9700F CPU and 16 GB of RAM, while the software environment was based on Python 3.9. In this study, the number of oracle nodes was set to \(N=21\), with each node collecting data from \(M=5\) sources, reaching consensus based on a threshold of \(t=11\). To simulate data heterogeneity and latency fluctuations in real-world scenarios, node response times were randomly generated using a Gaussian distribution, with slight uniform perturbations applied to enhance data representativeness. The state of the real-time data sources varied at a frequency of \(f\). Comparative methods included common aggregation strategies such as median \cite{SchellingCoin,Oracul,tellor,Compound}, mode, and majority voting \cite{dong2023daon}. Detailed experimental parameters are presented in Table \ref{tab:setting}.

\begin{figure*}[h]
\centering
\subfloat[Gaussian distribution ($f=5$ Hz)]{\includegraphics[width=0.3\linewidth]{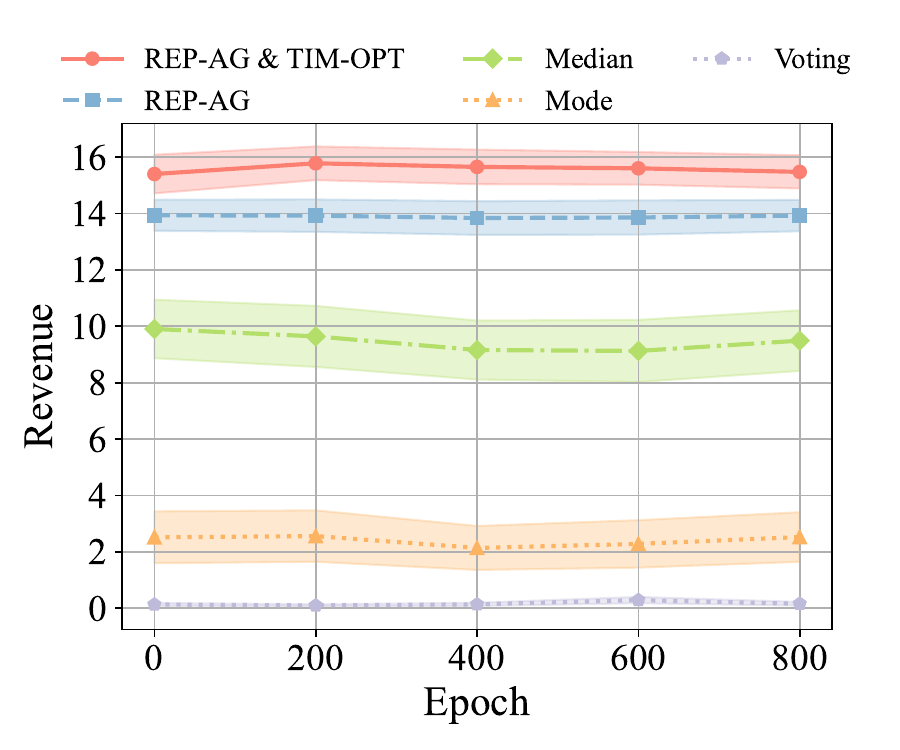}%
\label{fig:benefit_gaussain}}
\hfil
\subfloat[Uniform distribution ($f=5$ Hz)]{\includegraphics[width=0.3\linewidth]{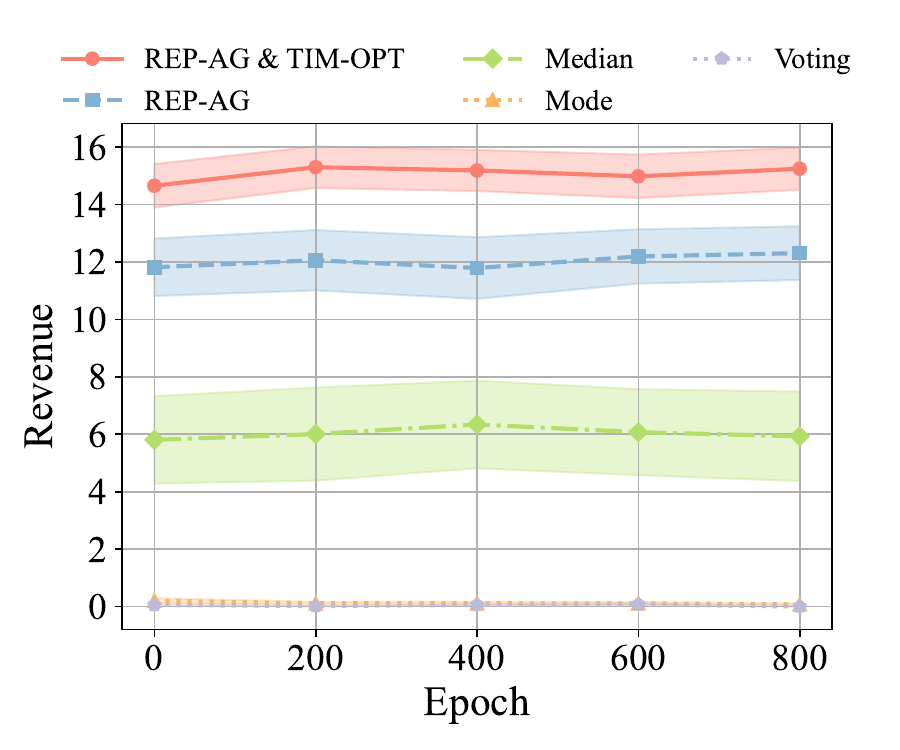}%
\label{fig:benefit_random}}
\hfil
\subfloat[Gaussian distribution (First 50 Tasks)]{\includegraphics[width=0.3\linewidth]{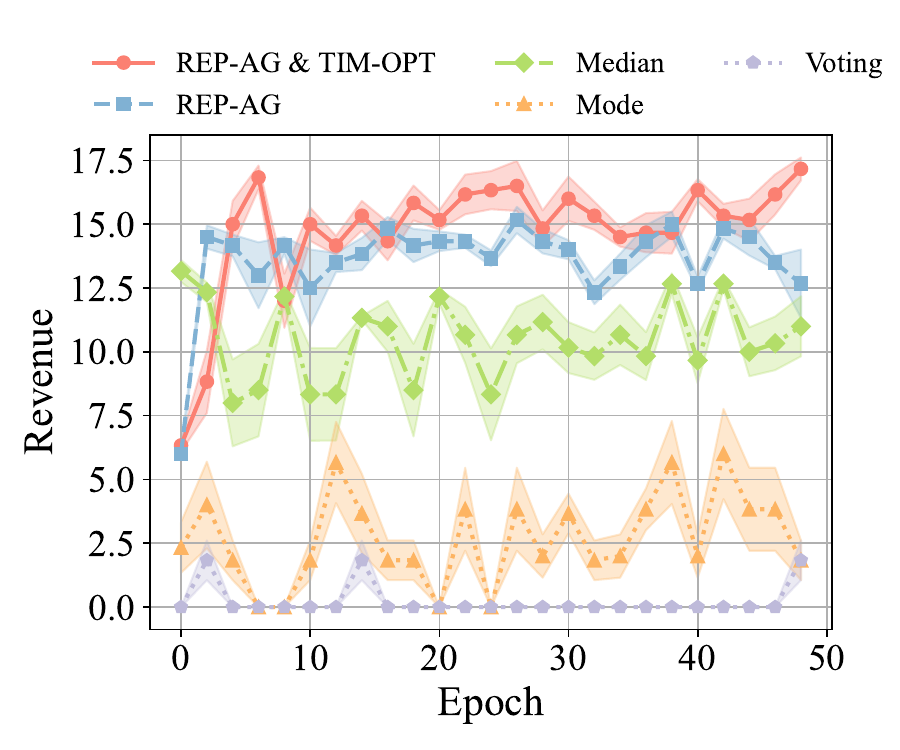}%
\label{fig:benefit_gaussain_50}}
\caption{Benefit Curves}
\label{fig:benefit}
\end{figure*}

\subsection{Consensus Success Rate}
Figure \ref{fig:all_gaussian} illustrates the consensus success rates of nodes employing various data aggregation methods to select data representatives under a Gaussian network latency distribution, based on the execution results of 1,000 tasks. We compared the consensus success rates of the median, mode, majority voting, and REP-AG, and analyzed the results after applying TIM-OPT. The experimental results indicate that REP-AG is more effective than traditional aggregation methods in assisting nodes to discover data representatives, resulting in a high level of consistency among node data, thereby significantly enhancing the consensus success rate of the threshold signature algorithm. Under varying data source change frequencies \(f\), REP-AG achieved an approximate 56.6\% improvement over the optimal baseline. After the application of TIM-OPT, the average consensus success rate of all methods increased by about 32.9\%. These findings validate the significant effectiveness of both REP-AG and TIM-OPT in enhancing the consensus success rate of threshold signatures.

Figure \ref{fig:all_random} presents the consensus success rates of nodes utilizing different data aggregation methods to select data representatives under a uniform network latency distribution, based on the execution results of 1,000 tasks. The experimental results are consistent with those shown in Figure \ref{fig:all_gaussian}. Although the difficulty of nodes obtaining identical data increases with the change frequency \(f\), leading to a downward trend in consensus success rates, our proposed methods still demonstrate a clear advantage. In this environment, the effectiveness of TIM-OPT is even more pronounced, yielding an average improvement of nearly 65.8\%. This further underscores the performance advantages and universality of the proposed approaches.

Additionally, we identified the fundamental reason behind the low success rates of aggregation schemes based on redundant data (such as mode and voting). For data sources with high-frequency changes, these methods struggle to establish a mode, and the availability of data meeting the majority voting requirement \(\left \lfloor \frac{N}{2} \right \rfloor + 1\) becomes increasingly scarce. Consequently, these strategies ultimately degenerate into random selection, which, even with the introduction of TIM-OPT, fails to significantly enhance their consensus success rates.

\begin{figure}[h]
\centering
{\includegraphics[width=0.8\linewidth]{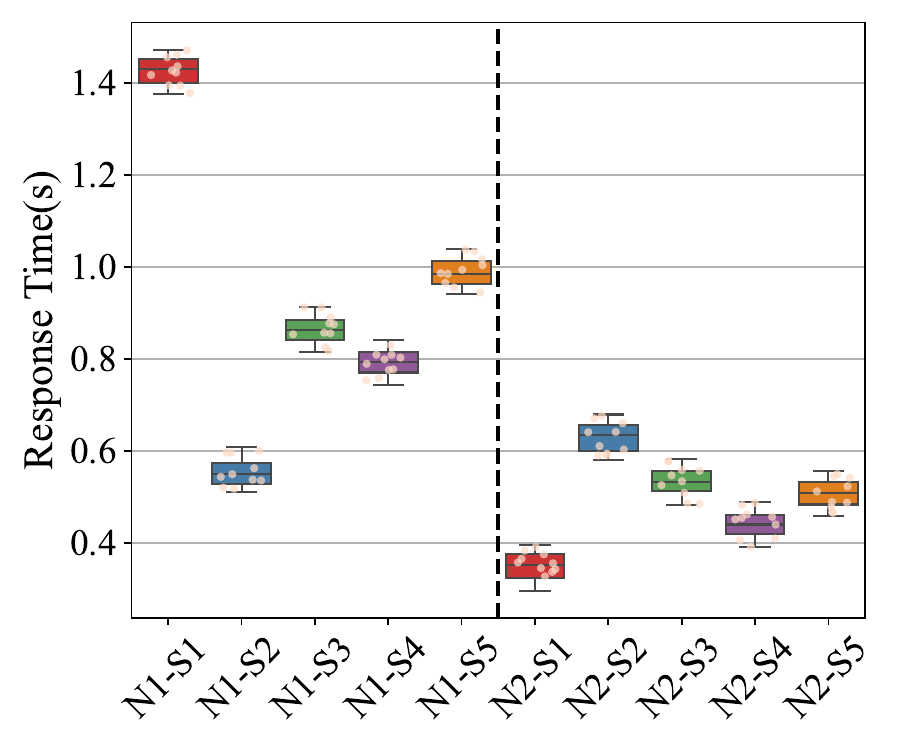}%
}
\caption{Response timing distribution for two randomly sampled nodes.}
\label{fig:time1}
\end{figure}

\begin{figure}[h]
\centering
{\includegraphics[width=0.8\linewidth]{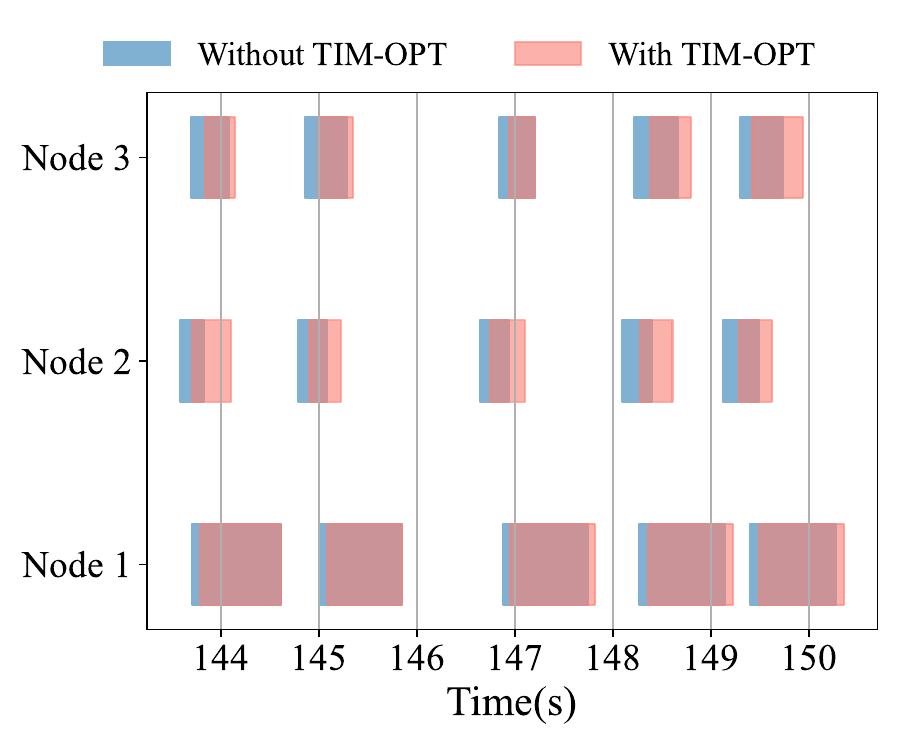}%
}
\caption{Changes in response timing for three randomly sampled nodes after the application of TIM-OPT.}
\label{fig:time2}
\end{figure}

\subsection{Benefit Curves}
Figures \ref{fig:benefit_gaussain} to \ref{fig:benefit_random} illustrate the benefit curves of the nodes under different network latency distributions in various aggregation schemes. The results indicate that the total benefit achieved by the proposed methods approaches approximately 16, suggesting that nearly 16 nodes return consistent data in each task. This further validates that the proposed schemes effectively enhance the consistency of the data representatives selected by nodes compared to traditional data aggregation methods, thereby improving node benefits.

Figure \ref{fig:benefit_gaussain_50} displays the benefit curves for the first 50 tasks. REP-AG and TIM-OPT enable most nodes to identify optimal strategies with only a few attempts during the early stages of the tasks, leading to rapid benefit enhancement. However, the final benefit does not reach \(N=21\), indicating that for a small number of nodes with significant deviations, even adjustments to their access times through TIM-OPT are insufficient to align them with the access timings of the majority. To address this issue, we plan to propose potential solutions in future work.

\subsection{Response Timing Analysis}
Figure \ref{fig:time1} illustrates the response timing distribution of two randomly sampled nodes for $M=5$ data sources under a Gaussian delay distribution. Taking a frequency of $f=5$, the data state can change every 0.2 seconds. Taking the mode strategy as an example, the modes for nodes $N_1$ and $N_2$ typically fall within different time intervals. It demonstrates the failure of traditional strategies such as the mode: in a heterogeneous environment, the data distribution acquired by nodes differs, resulting in inconsistencies between their modes, which may even lead to the absence of a mode.

Figure \ref{fig:time2} shows the effectiveness of the proposed TIM-OPT. We randomly sampled the response time distributions of three nodes in five tasks. The blue area indicates the time intervals without TIM-OPT, while the red area represents the intervals after applying TIM-OPT. TIM-OPT exerts a greater adjustment effect on nodes with smaller response time intervals (e.g., Node2 and Node3), while adjustments to Node1 are relatively minimal. Ultimately, the response time intervals of the three nodes become more aligned, thereby increasing the likelihood of data consistency. This further validates the effectiveness of TIM-OPT.

\subsection{Strategy Analysis}

\begin{figure}[h]
\centering
{\includegraphics[width=0.8\linewidth]{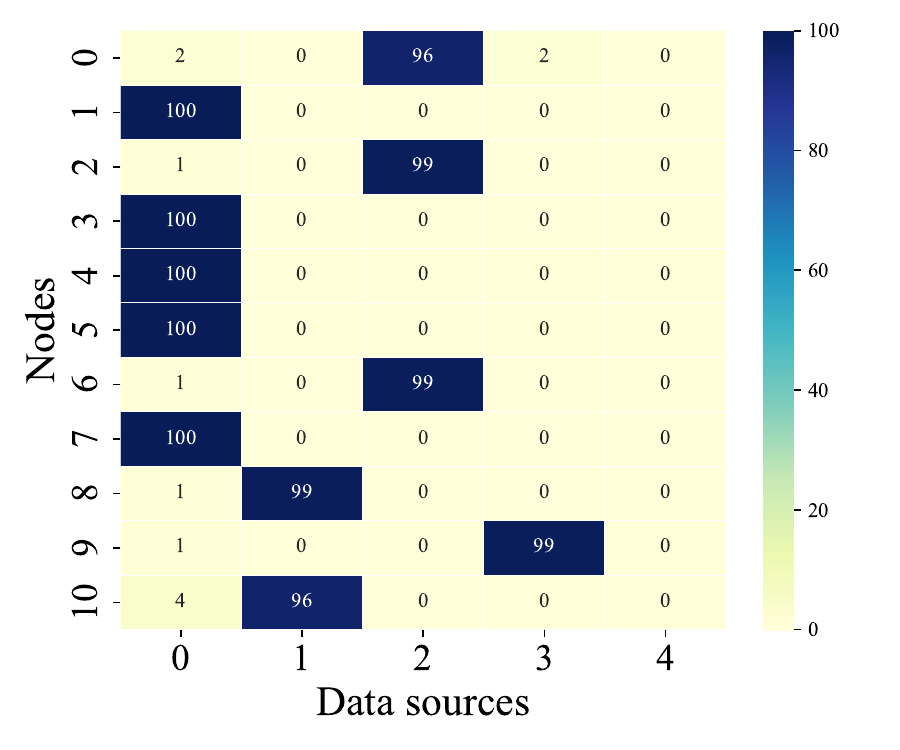}%
}
\caption{Data representative selection details of REP-AG when \( f = 5 \) Hz.}
\label{fig:analysis_5}
\end{figure}

\begin{figure}[h]
\centering
{\includegraphics[width=0.8\linewidth]{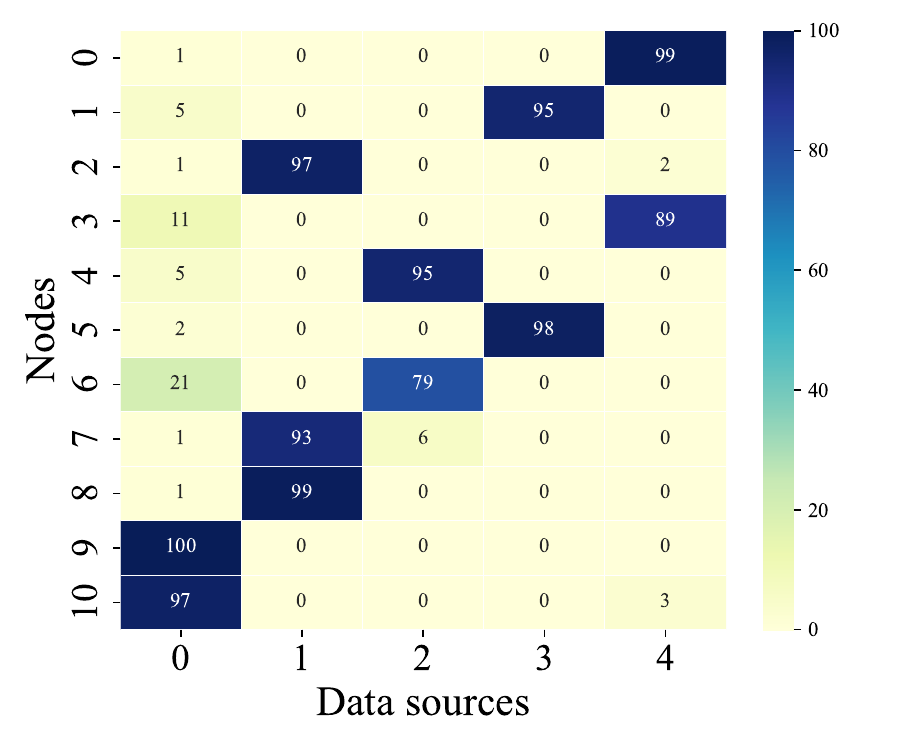}%
}
\caption{Data representative selection details of REP-AG when \( f = 10 \) Hz.}
\label{fig:analysis_10}
\end{figure}

Figures \ref{fig:analysis_5} and \ref{fig:analysis_10} illustrate the details of REP-AG data representative selection at varying frequencies \( f \). When \( f = 5 \), nodes can determine their optimal data representatives with minimal exploration. However, at \( f = 10 \), the selection of data representatives becomes more dispersed between nodes, complicating the decision-making process. For example, Node 6 engaged in extensive trials when selecting data from data sources 0 and 2. Nevertheless, the REP-AG strategy ultimately encourages most nodes to stabilize their strategies, leading to a gradual convergence towards an approximate equilibrium state in overall revenue levels. This indicates that while some nodes may still have the incentive to adjust their strategies, the majority have essentially converged on effective data representative selections after iterations of the REP-AG strategy.

\begin{figure*}[h]
\centering
{\includegraphics[width=1\linewidth]{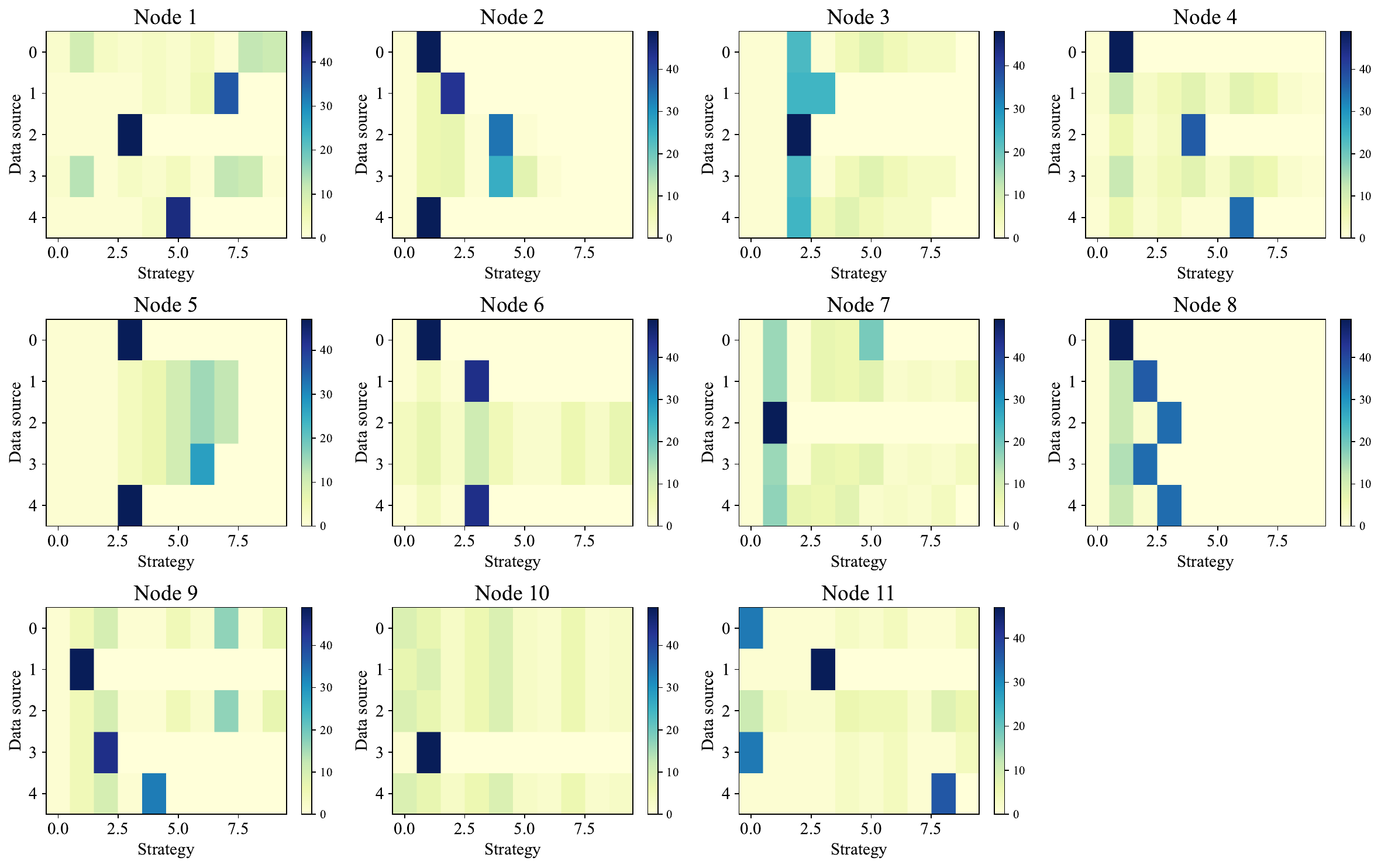}%
}
\caption{Strategy selection details of TIM-OPT when \( f = 5 \) Hz.}
\label{fig:analysis_detail}
\end{figure*}

Figure \ref{fig:analysis_detail} provides information on the strategy choices made by TIM-OPT on 50 tasks. For certain data sources, as indicated by the darker regions, TIM-OPT effectively identifies optimal access timing strategies, enhancing the consistency of data across nodes. However, for specific strategies, such as those involving data sources 0-2 in Node 10, the access times may differ significantly from those of other nodes, complicating the learning process for optimal strategies. Overall, TIM-OPT assists the majority of nodes in converging to optimal response timing strategies after 50 task interactions, resulting in minimal motivation to further adjust their strategies and achieve a state of near equilibrium.

\begin{figure*}[htp!]
\centering
\subfloat[Gaussian distribution \( f = 5 \, \text{Hz} \)]{\includegraphics[width=0.33\linewidth]{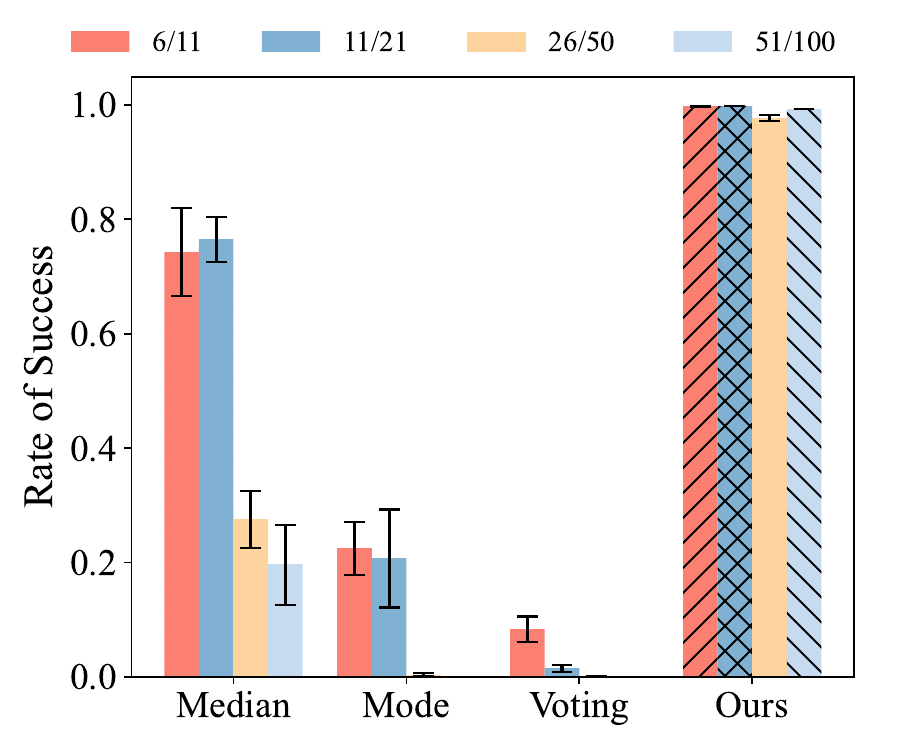}%
\label{fig:robustness_gaussian}}
\hfil
\subfloat[Random distribution \( f = 5 \, \text{Hz} \)]{\includegraphics[width=0.33\linewidth]{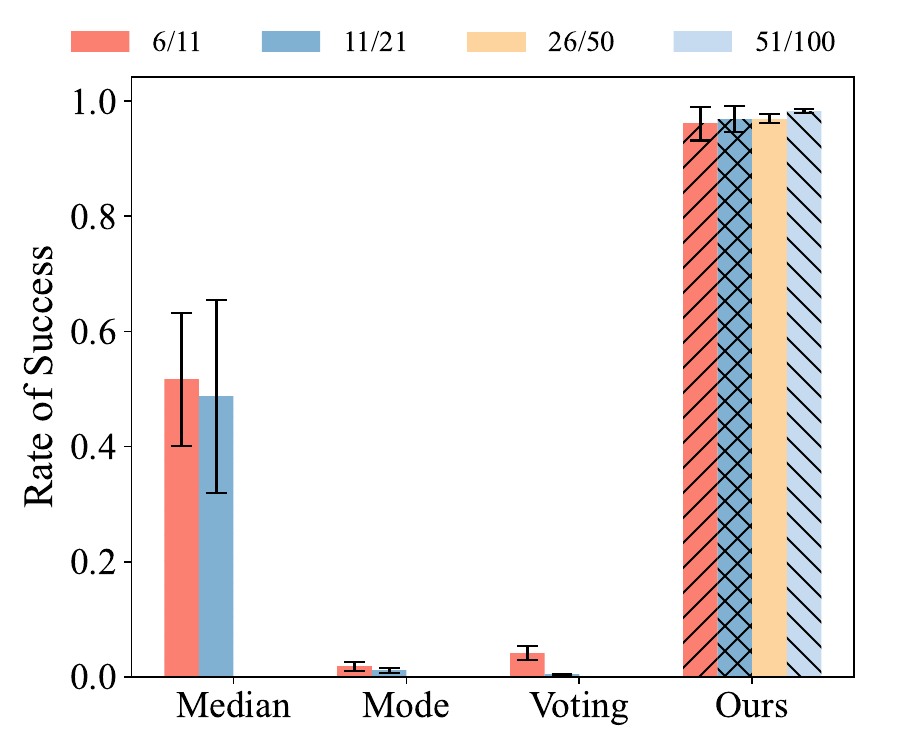}%
\label{fig:robustness_random}}
\hfil
\subfloat[Different threshold requirements]{\includegraphics[width=0.33\linewidth]{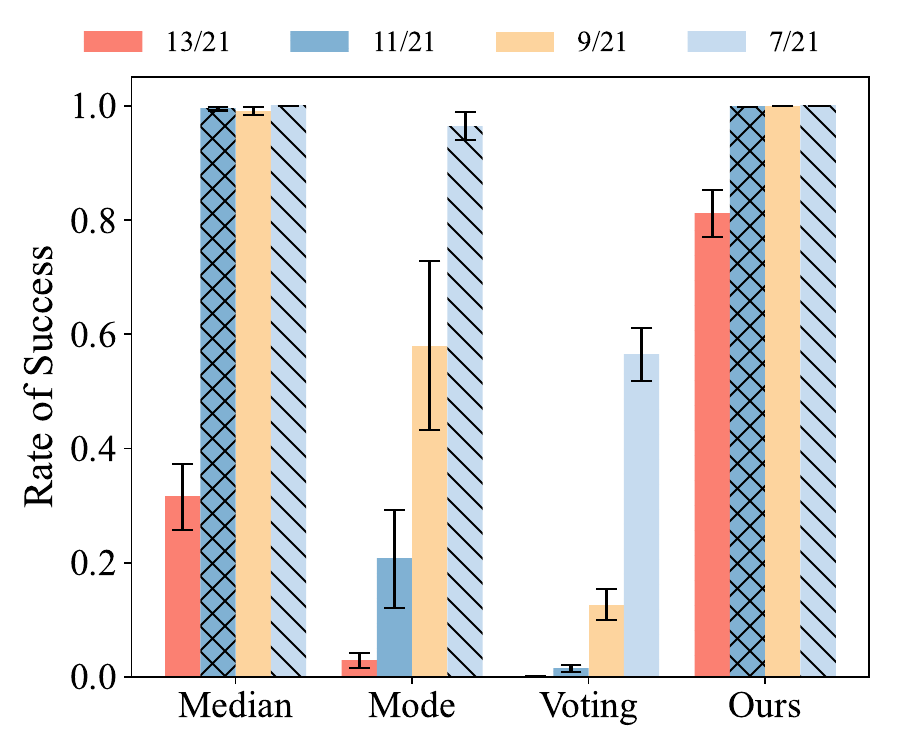}%
\label{fig:robustness_t}}
\caption{Robustness analysis.}
\label{fig:robustness}
\end{figure*}

\subsection{Robustness}
Figures \ref{fig:robustness_gaussian} and \ref{fig:robustness_random} illustrate the impact of increasing the number of nodes \( N \) on consensus success rates under different network latency distributions. The proposed scheme maintains effective performance in various environments. In particular, as shown in Fig. \ref{fig:robustness_random}, the overall consensus success rate tends to decrease with increasing \( N \). This observation aligns with our analysis in the \ref{random_success}, which explains the low consensus success rates observed in traditional schemes under real-time data sources. In contrast, our approach fosters the emergence of an approximate equilibrium state among nodes through belief iteration under incomplete information, demonstrating strong robustness independent of \( N \).

Figure \ref{fig:robustness_t} examines the effect of varying threshold requirements \( t \) on consensus success rates. As expected, this finding corroborates the conclusions drawn in the \ref{random_success}. Lowering \( t \) facilitates consensus but also directly impacts the security of threshold signatures; thus, it is essential to ensure \( t > \frac{N}{2} \) for safety. Furthermore, our proposed scheme effectively maintains consensus success rates even when \( t > 11 \), further supporting its robustness.

\section{Conclusion}
\label{conclusion}
This paper presents an innovative dual strategy that effectively enhances the consensus success rate of threshold signature oracles in acquiring real-time data. To improve the consistency of aggregation results, we have developed a novel data aggregation method, REP-AG, which utilizes Bayesian game theory to select aggregated data that is most likely similar to that of other nodes. Furthermore, we introduce the Time of Access Optimization Strategy (TIM-OPT), which adjusts the timing of nodes' access to data sources, significantly increasing data centrality. Experimental results validate the effectiveness of both proposed strategies.

In the future, we aim to enhance data sharing among nodes while ensuring system security, thereby improving the consensus success rate of threshold signatures. Additionally, we plan to extend our research to encompass a broader range of real-time data consensus scenarios, not only limited to blockchain oracles but also including fast consensus algorithms and other areas. Ultimately, we hope to provide new insights into the application and development of blockchain technology by exploring additional scenarios in real-time data applications.

\appendix
\section{Consensus success rate analysis of threshold signature under uniform delay distribution}
\label{random_success}

First, as a single node accesses data from \( M \) data sources characterized by a uniform response time distribution \( U(a,b) \), it randomly selects one data point as its representative to participate in the consensus of the threshold signature. To simplify the analysis, we assume that the data representative selection process is random. Based on this, we perform a formal analysis of the probability that \( N \) nodes obtain at least \( t \) identical data points while accessing data sources with a change frequency of \( f \) Hz.

\subsection{Parameter Definitions}

\begin{itemize}
    \item \textbf{Number of Nodes \( N \)}: The number of nodes accessing the data sources.
    \item \textbf{Change Frequency \( f \)}: The frequency at which data sources change per second (that is, \( f \) changes per second).
    \item \textbf{Response Time Distribution}: Each node's response time follows a uniform distribution \( U(a, b) \).
    \item \textbf{Expected Number of Identical Data \( t \)}: The desired number of identical data points.
    \item \textbf{Total Duration \( T \)}: \( T = b - a \).
    \item \textbf{Length of Each Data Return Interval}: The time length during which a data source returns the same data upon changing is \( \frac{1}{f} \).
    \item \textbf{Number of Possible Different Data Return Intervals}: There will be \( (b - a) f \) possible intervals for returning different data.
\end{itemize}

\subsection{Probability Calculations}

\paragraph{Node Distribution in Each Interval}

We focus on the distribution of \( N \) nodes' response times across \( (b - a) f \) time intervals. Our goal is to calculate the probability that at least \( t \) nodes have response times falling within a specific time interval \( i \).

\paragraph{Probability of Node Overlap in a Single Interval}

Let the event that at least \( t \) nodes overlap in the interval \( i \) be denoted as \( E_i \).

\begin{itemize}
    \item \textbf{Probability of Each Node Falling in Interval \( i \)}:
    \begin{eqnarray}
    P(\text{node in interval } i) = \frac{\frac{1}{f}}{b - a} = \frac{1}{f(b - a)}
    \end{eqnarray}
    
    \item \textbf{Probability of \( k \) Nodes in Interval \( i \)}: According to the binomial distribution, the probability of \( k \) nodes falling in interval \( i \) is:
    \begin{eqnarray}
    P(k \text{ nodes in } E_i) = \binom{N}{k} p^k (1 - p)^{N - k}
    \end{eqnarray}
        where \( p = \frac{1}{f(b - a)} \) is the probability of a single node falling in the interval \( i \).
    \end{itemize}

\paragraph{Probability of At Least \( t \) Nodes}

Thus, the probability of having at least \( t \) nodes in interval \( i \) is given by:
\begin{eqnarray}
P(E_i) = \sum_{k=t}^{N} \binom{N}{k} p^k (1 - p)^{N - k}
\end{eqnarray}

\paragraph{Total Probability}

Considering all \( (b - a) f \) intervals, the total probability of obtaining the same data across the entire system is:
\begin{eqnarray}
P(\text{at least } t \text{ same data}) = 1 - \prod_{i=1}^{(b - a)f} \left(1 - P(E_i)\right)
\end{eqnarray}

\paragraph{Consensus Success Rate}

\begin{eqnarray}
P_t^N = 1 - \prod_{i=1}^{(b - a)f} \left(1 - \sum_{k=t}^{N} \binom{N}{k} p^k (1 - p)^{N - k}\right)
\end{eqnarray}

Thus, the consensus success rate \( P_t^N \) reveals that it is closely related not only to the nodes' response time distribution \( (a,b) \) and the data source change frequency \( f \), but also to the total number of nodes \( N \) and the threshold \( t \). Table \ref{tab:success_rate} illustrates the intuitive impact of varying \( N \) and \( t \) on the consensus success rate. It is evident that under the aforementioned assumptions, the consensus success rate is negatively correlated with \( t \). Even when maintaining the same security conditions, specifically \( t=\left \lfloor \frac{N}{2} \right \rfloor +1 \), the consensus success rate declines as \( N \) increases.


\begin{table}[h]
\centering
\caption{Theoretical consensus success rates for threshold signatures}
\label{tab:success_rate}
\resizebox{\linewidth}{!}{%
\begin{tabular}{ccc}
\hline
Number of Nodes & Threshold & Consensus Success Rate \\ \hline
11 & 2 & 99.36\% \\
11 & 4 & 45.26\% \\
\rowcolor{lightgray} 11 & 6 & 2.73\% \\
\rowcolor{lightgray} 21 & 11 & 0.11\% \\
\rowcolor{lightgray} 50 & 26 & 6.51$\times 10^{-6}$\% \\
\rowcolor{lightgray} 100 & 51 & 1.99$\times 10^{-12}$\% \\ \hline
\end{tabular}
}
\end{table}

\section*{CRediT authorship contribution statement}
Youquan Xian: Conceptualization, Formal analysis, Investigation, Methodology, Software, Validation, Visualization, Writing – original draft. Xueying Zeng: Conceptualization, Investigation, Writing– Review \& Editing. Chunpei Li, Dongcheng Li, and Peng Wang: Supervision, Review \& Editing. Peng Liu, and Xianxian Li: Supervision, Review, Editing \& Funding acquisition.

\section*{Declaration of competing interest}
The authors declared that they have no conflicts of interest to work.

\section*{Data availability}
Data will be made available on request.

\section*{Acknowledgments}
This work was supported in part by the Guangxi Science and Technology Major Project (No. AA22068070), the National Natural Science Foundation of China (Nos. 62166004, U21A20474), the Basic Ability Enhancement Program for Young and Middle-aged Teachers of Guangxi (No. 2023KY0062), Innovation Project of Guangxi Graduate Education (Nos. XYCBZ2024025, XYCSR2024098).

 \bibliographystyle{elsarticle-num} 
 \bibliography{myref}


\subsection*{  }
\setlength\intextsep{0pt} 
\begin{wrapfigure}{l}{0.12\textwidth}
    \centering
    \includegraphics[width=0.15\textwidth]{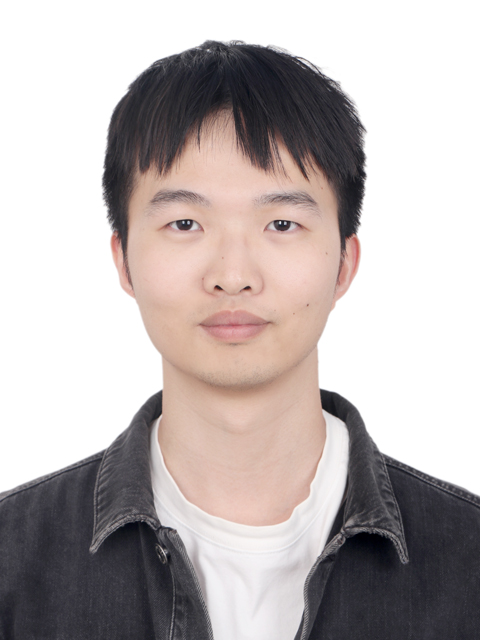}
\end{wrapfigure}
\noindent \textbf{Youquan Xian} received his master's degree from Guangxi Normal University in 2024. He has published multiple papers in journals and conferences such as IEEE Transactions on Network and Service Management, IEICE Transactions on Communications, IEEE SMC 2024, and WASA 2024. His main research directions include blockchain and federated learning.

\subsection*{  }
\setlength\intextsep{0pt}
\begin{wrapfigure}{l}{0.12\textwidth}
    \centering
    \includegraphics[width=0.15\textwidth]{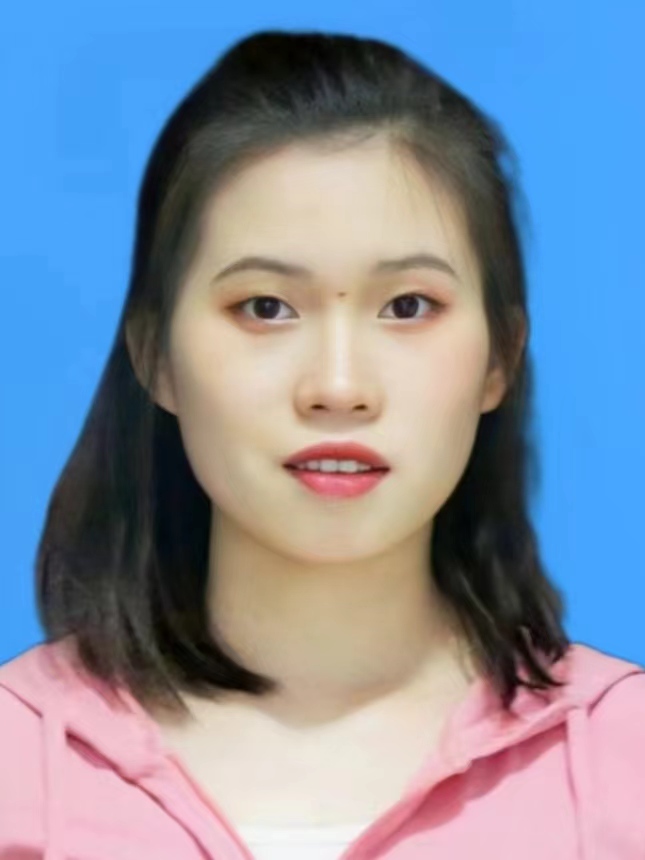}
\end{wrapfigure}
\noindent \textbf{Xueying Zeng} received her bachelor's degree from Guangxi Science and Technology Normal University in 2022. She is currently pursuing her master's degree from Guangxi Normal University. Her research interests are blockchain, crowdsourcing.

\subsection*{  }
\setlength\intextsep{0pt}
\begin{wrapfigure}{l}{0.12\textwidth}
    \centering
    \includegraphics[width=0.15\textwidth]{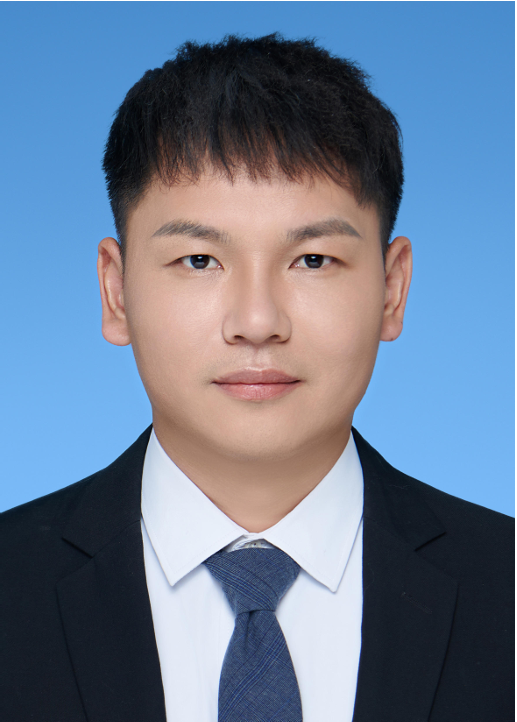}
\end{wrapfigure}
\noindent \textbf{Chunpei Li} received his Ph.D. from the School of
 Computer Science and Engineering at Guangxi Normal University in 2024. He is currently conducting
 postdoctoral research at the Ministry of Education
 Key Laboratory of Educational Blockchain and Intelligent Technology at Guangxi Normal University.
 His research interests include blockchain, artificial
 intelligence, and information security.

\subsection*{  }
\setlength\intextsep{0pt}
\begin{wrapfigure}{l}{0.12\textwidth}
    \centering
    \includegraphics[width=0.15\textwidth]{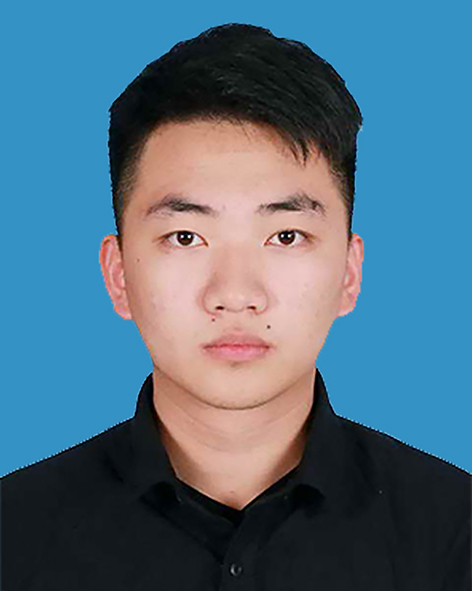}
\end{wrapfigure}
\noindent \textbf{Dongcheng Li} received his master's degree in sofrware engineering from  Guangxi normal university. He  is currently working at the department of 
Computer Science and Engineering of Guangxi normal university, China. His main research interests include blockchain, data security and recommendation system.

\subsection*{  }
\setlength\intextsep{0pt}
\begin{wrapfigure}{l}{0.12\textwidth}
    \centering
    \includegraphics[width=0.15\textwidth]{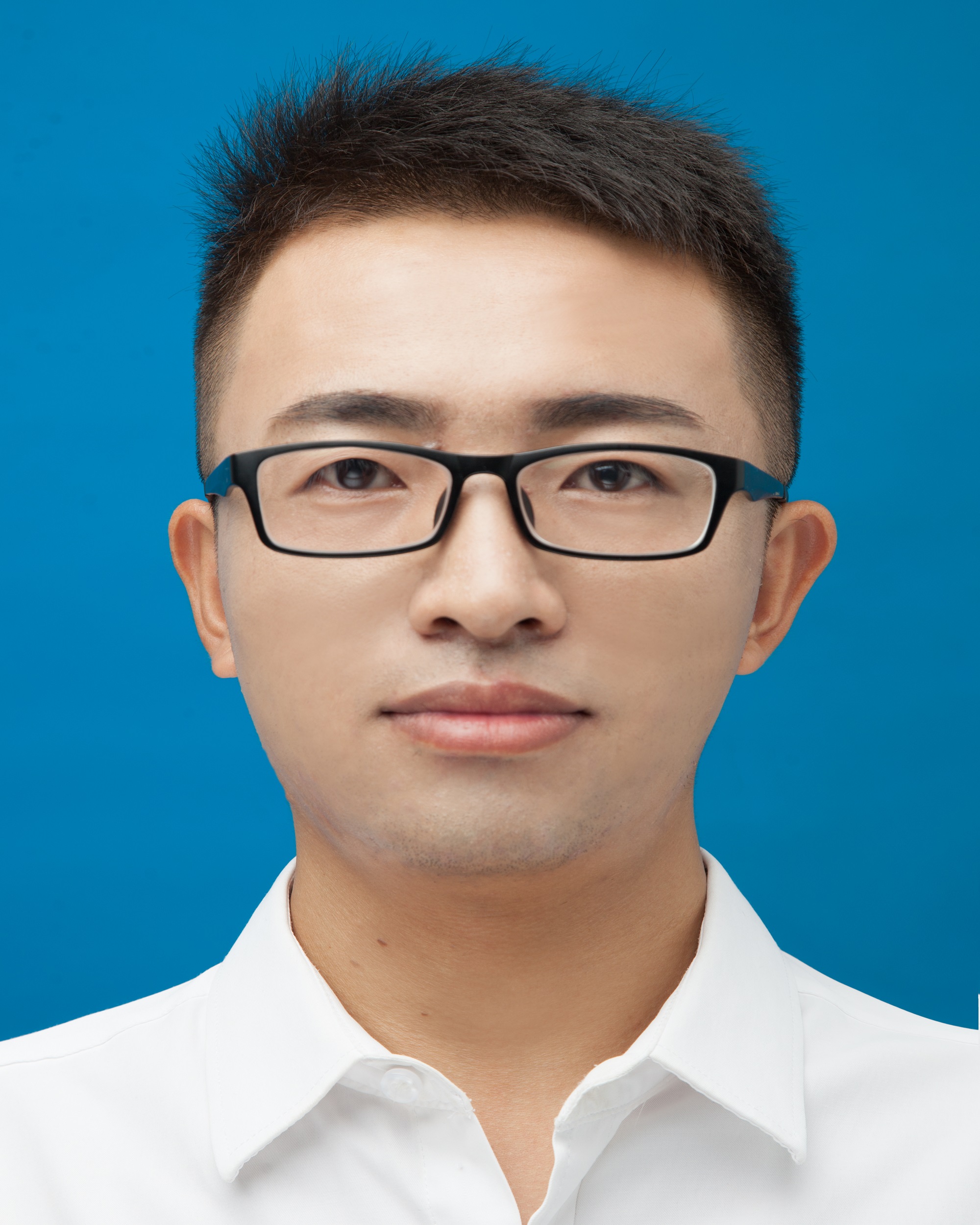}
\end{wrapfigure}
\noindent \textbf{Peng Wang} received his master's degree from Guilin University of Technology in 2018. He is currently working toward a doctor's degree at Guangxi Normal University. His research interests include blockchain, data fusion, and data security.

\subsection*{  }
\setlength\intextsep{0pt}
\begin{wrapfigure}{l}{0.12\textwidth}
    \centering
    \includegraphics[width=0.15\textwidth]{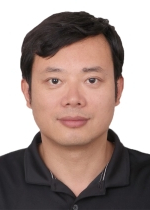}
\end{wrapfigure}
\noindent \textbf{Peng Liu} received his Ph.D. degree in 2017 from Beihang University, China. He began his academic career as an assistant professor at Guangxi Normal University in 2007 and was promoted to full professor in 2022. His current research interests are focused on federated learning and blockchain.

\subsection*{  }
\setlength\intextsep{0pt}
\begin{wrapfigure}{l}{0.12\textwidth}
    \centering
    \includegraphics[width=0.15\textwidth]{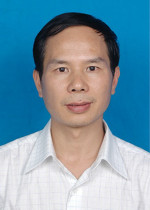}
\end{wrapfigure}
\noindent \textbf{Xianxian Li} received his Ph.D. degree from the School of Computer Science and  Engineering,  Beihang University, Beijing, China, in 2002. He worked as a professor at Beihang University during 2003-2010. He is currently a professor with  the School of Computer Science and Engineering, Guangxi Normal University, Guilin, China. His research interest includes information security.

\end{document}